\documentclass[%
 aip,
 amsmath,amssymb,
 reprint,%
]{revtex4-1}
\usepackage{graphicx}
\usepackage{subfigure}
\usepackage{dcolumn}
\usepackage{bm}
\usepackage{color}
\usepackage[T1]{fontenc}
\usepackage{mathptmx}
\usepackage{multirow}
\usepackage{array}
\usepackage{float}
\usepackage{url}
\usepackage{CJKutf8}
\usepackage{pifont}
\usepackage{epstopdf, epsfig}
\usepackage{graphicx}
\usepackage{dcolumn}
\usepackage{longtable,booktabs}
\usepackage{upgreek }
\usepackage{textcomp}
\usepackage{ wasysym }
\usepackage{titlesec}
\usepackage{float}
\usepackage{color,xcolor}
\usepackage{graphicx,times}
\usepackage{natbib}
\usepackage{amssymb,amsmath}
\usepackage{epstopdf}
\usepackage{graphicx}
\usepackage{subfigure}
\usepackage{makecell}
\usepackage{multirow}
\usepackage[
    colorlinks=true,
    linkcolor=blue,
    urlcolor=blue,
    citecolor=blue
]{hyperref}
\usepackage{subfigure}
\usepackage{makecell}
\usepackage{multirow}

\usepackage{titlesec}
\titleclass{\subsubsubsection}{straight}[\subsection]
\newcounter{subsubsubsection}[subsubsection]
\renewcommand\thesubsubsubsection{\thesubsubsection.\alph{subsubsubsection}}
\titleformat{\subsubsubsection}
 {\normalfont\normalsize\bfseries}{\thesubsubsubsection}{1em}{}
\titlespacing*{\subsubsubsection}
{0pt}{3.25ex plus 1ex minus .2ex}{1.5ex plus .2ex}

\renewcommand\arraystretch{1.5}

\begin{document}
\preprint{AIP/123-QED}

\title{Droplet coalescence kinetics: Coalescence mechanisms and thermodynamic non-equilibrium effects under isothermal and non-isothermal conditions}

\author{Guanglan Sun \begin{CJK*}{UTF8}{gbsn} (孙光兰) \end{CJK*}}
 \affiliation{Hebei Key Laboratory of Trans-Media Aerial Underwater Vehicle,
 School of Liberal Arts and Sciences,
North China Institute of Aerospace Engineering, Langfang 065000, China}
\author{Yanbiao Gan \begin{CJK*}{UTF8}{gbsn} (甘延标) \end{CJK*}}
 \thanks{Corresponding author: gan@nciae.edu.cn}
  \affiliation{Hebei Key Laboratory of Trans-Media Aerial Underwater Vehicle,
  School of Liberal Arts and Sciences,
North China Institute of Aerospace Engineering, Langfang 065000, China}
\author{Bin Yang \begin{CJK*}{UTF8}{gbsn} (杨斌) \end{CJK*}}
  \affiliation{School of Energy and Safety Engineering, Tianjin Chengjian University, Tianjin 300384, China}
\author{Aiguo Xu \begin{CJK*}{UTF8}{gbsn} (许爱国) \end{CJK*}}
\affiliation{National Key Laboratory of Computational Physics, Institute of Applied Physics and Computational Mathematics, P. O. Box 8009-26, Beijing 100088, P.R.China}
\affiliation{National Key Laboratory of Shock Wave and Detonation Physics, Mianyang 621999, China}
\affiliation{HEDPS, Center for Applied Physics and Technology, and College of Engineering, Peking University, Beijing 100871, China}
\affiliation{State Key Laboratory of Explosion Science and Safety Protection, Beijing Institute of Technology, Beijing 100081, China}
\author{Zhipeng Liu \begin{CJK*}{UTF8}{gbsn} (刘枝朋) \end{CJK*}}
\thanks{Corresponding author: zhipengliu@tcu.edu.cn}
\affiliation{School of Science, Tianjin Chengjian University, Tianjin 300384, China}

\date{\today}

\begin{abstract}
Droplet coalescence is a multiscale phenomenon driven by multiple thermodynamic non-equilibrium (TNE) effects. This study investigates the coalescence mechanisms and the complex interplay between various TNE effects in both isothermal and non-isothermal cases kinetically.
The main findings include:
(1) Coalescence initiation and cut-through mechanisms: In the non-isothermal case, latent heat release results in a temperature rise that slightly increases the surface tension gradient (driving force) near the contact point, while substantially enhancing the pressure gradient (resistance). This leads to a significantly prolonged initiation of coalescence compared to the isothermal case. The additional temperature field effects introduce greater spatial complexity to both the flow field and morphology. For the cut-through mechanism, in both cases, pressure extends the liquid-vapor interface in opposite directions, promoting liquid bridge growth.
(2) TNE effects: The latent heat-induced temperature rise in the thermal case considerably dampens the intensity of TNE effects. Before and after droplet contact, TNE quantities driven by the temperature gradient ($\bm \Delta^{\ast}_{3,1}$ and $\bm \Delta^{\ast}_{3}$) and those driven by the velocity gradient ($\bm \Delta^{\ast}_{2}$ and $\bm \Delta^{\ast}_{4,2}$), alternate in dominating the coalescence process.
This alternating dominance result in a more complex spatiotemporal evolution of TNE effects compared to the isothermal case, where the TNE intensity is dominated by $\bm \Delta^{\ast}_{2}$ and $\bm \Delta^{\ast}_{4,2}$.
(3) Entropy production mechanisms: In the non-isothermal case, entropy production is contributed by both $\bm \Delta^{\ast}_2$ and $\bm \Delta^{\ast}_{3,1}$, with the former being the dominant contributor. The temperature field reduces the entropy production rate, but extends its duration, thereby increasing the total entropy production. The entropy production rates from $\bm \Delta^{\ast}_2$ and $\bm \Delta^{\ast}_{3,1}$ alternate before and after the droplet cut-through. 
Further investigation into effects of the initial droplet distance $r_d$ revealed that it increases droplet cut-through time $t_c$, the time when total TNE reaches its maximum $t_{D\rm max}$, and the time when total entropy production rate reaches its maximum $t_{\dot S \rm pr max}$ in a power-law manner: $(t_c,t_{D\rm max},t_{\dot S \rm pr max}) \sim {r_d}^{\alpha}$.
This research provides kinetic insights into droplet coalescence, offering mesoscopic support for dynamic, cross-scale regulation and multifunctional integration of coalescence processes in industrial applications.
\end{abstract}
\maketitle
\section{\label{sec:level1} Introduction}

Droplet coalescence is a widespread phenomenon in fields such as nature, biomedical sciences, microfluidics, engine combustion, multi-medium vehicles, and photolithography\cite{RN709, RN732,RN288, RN725,RN743,RN651,RN652,RN252}.
In the field of microfluidics, droplet-based technologies involving deformation, fragmentation, coalescence, and collision, hold great potential for applications in medical diagnostics, food science, drug development, and artificial biomimetic systems manufacturing\cite{RN709, RN732,RN252}.
Droplets, serve as multifunctional carriers for controlled reactions, mixing, and analysis, play a key role in these processes\cite{RN709, RN732,RN874}.
In combustion systems with dense spray devices, intense interactions between droplets-including atomization, collision, rupture, coalescence, and oscillation-greatly affect fuel heat and mass transfer, combustion efficiency, and combustion stability\cite{RN288, RN725}.
In explosion dynamics, studying droplet deformation, collision, and coalescence under shock waves at various Mach numbers provides crucial insights for mitigating shockwave energy and reducing shock damage\cite{RN743}.

The physical mechanisms and evolutionary patterns of droplet collision, coalescence, and fragmentation have been extensively studied under isothermal conditions. Surface tension is the primary driving force for the spontaneous coalescence of static droplets, which reduces the system's free energy by minimizing the total inter-droplet surface area. In the initial stage of static droplet coalescence, the growth of the liquid bridge radius $r_b$ over time is directly related to the system's viscosity\cite{RN296, RN298, RN227, RN452, RN297, RN420, RN757, RN758}. The shape of the droplet after collision, coalescence, and fragmentation is closely linked to characteristic parameters such as the Bond number (Bo), Weber number (We), Ohnesorge number (Oh), and the initial radius ratio of the droplets\cite{RN320, RN759, RN760, RN762, RN764, RN761, RN766, RN765, RN763, RN767, RN43, RN739, RN662, RN677}. The number of droplets, the gas-liquid density ratio of the system, and the presence of chemical reactions between droplets also influence the morphological characteristics and evolutionary patterns of droplet coalescence\cite{RN326, RN770}.

The main difference between thermal and isothermal multiphase flow systems is the heat generation and conduction during phase separation, resulting in lower interfacial stresses and localized temperature changes due to latent heat release, which leads to new local mechanical equilibrium\cite{RN83}. Under non-isothermal conditions, the variety of driving mechanisms increases significantly, and the competition and coupling between these mechanisms intensify, resulting in a more complex droplet coalescence process.
{Specifically, because surface tension varies sensitively with temperature, a gradient forms across the droplet surface, propelling the droplet along the temperature gradient—a process known as the Marangoni effect.} This effect significantly influences coalescence dynamics\cite{RN663, RN769, RN771, RN773}. The local temperature difference resulting from the gradient also alters the liquid's viscosity, enhancing both heat and mass transfer, and the coupling between them.
These changes modify the internal convection patterns and interface behavior of the droplet, subsequently affecting its deformation velocity and coalescence dynamics\cite{RN770, RN663}.
Currently, the kinetic study of droplet collision, coalescence, and fragmentation under non-isothermal conditions remains limited.
Non-isothermal conditions, however, are prevalent in combustion systems\cite{RN288, RN725, RN770, RN663}, explosion dynamics\cite{RN743}, nuclear reactors\cite{RN772}, and microfluidics\cite{RN774, RN775}.
For instance, in microfluidics, the thermal Marangoni effect is often used to control the droplet coalescence process\cite{RN774, RN775}.
This highlights the importance of comparative studies on the driving mechanisms and underlying kinetic behaviors of droplet coalescence under isothermal and non-isothermal conditions.

Numerical simulations offer several advantages, such as high efficiency, low cost, strong repeatability, flexible control of variables, and the capacity to handle complex multi-scale systems. These qualities make them indispensable tools for studying droplet coalescence. Widely employed macroscopic simulation methods, based on the Navier-Stokes (NS) equations\cite{RN121,RN299,RN662, RN677, RN679, RN682,RN676,RN720, RN721, RN722}, and mesoscopic methods, such as the lattice Boltzmann Method (LBM)\cite{RN692, RN693, RN694, RN695, RN696, RN781, RN782, RN783, RN787, RN788,RN719,RN703,RN702, RN790, RN870, RN871} and discrete unified gas kinetic scheme \cite{RN881,RN882,RN883}, have made significant strides in advancing our understanding of this phenomenon. Despite significant progress, several aspects of droplet coalescence still require further development.

\emph{(i) Neglect of temperature field and compressibility effects}: Most studies currently rely on isothermal, incompressible models, neglecting the impact of temperature fields and compressible convection on flow behavior. As a result, these models fail to fully capture the intricate interactions between droplets in real-world multiphase flows.

\emph{(ii) Oversight of non-equilibrium behaviors and effects}: While most studies focus on the evolution of conserved quantities—such as mass, momentum, and energy—alongside slow-varying variables, they often overlook the evolution of non-conserved, fast-varying variables like thermodynamic non-equilibrium (TNE) and hydrodynamic non-equilibrium (HNE) behaviors.
 These non-equilibrium phenomena are closely tied to small-scale structures and fast modes, such as liquid bridges, droplet boundary layers, vortices, droplet fragments, splash phenomena, and surface fluctuations. In regions near these structures, discrete effects become pronounced, challenging the reliability of the methods discussed above. Furthermore, during rapid collisions and coalescence, the onset of fast modes impedes the system's return to thermodynamic equilibrium in a timely manner, leading to strong non-equilibrium effects that further undermine the validity of these approaches. On the mesoscopic scale, HNE and TNE arise from the deviation of the distribution function from equilibrium, highlighting system complexity. On the macroscopic scale, these non-equilibrium effects are intricately coupled with macroscopic quantities and exert a fundamental influence on the overall behavior of the system.

\emph{(iii) Limitations in data analysis and information extraction}: In the droplet coalescence process, various driving mechanisms—including TNE effects, surface tension, temperature gradients, and velocity gradients—interact to produce complex, multi-scale spatiotemporal structures. To achieve a deeper understanding of these processes, a multi-perspective analysis is essential.

\emph{(iv) Lack of systematic comparison of entropy production mechanisms}: Entropy production mechanisms are pivotal in describing the thermodynamic behavior during droplet coalescence, particularly with regard to energy conversion and dissipation in non-equilibrium states. Despite their importance, existing studies lack a systematic comparison of entropy production mechanisms, especially in terms of the contributions of different driving forces—such as surface tension, heat flux, and viscous stress—to the entropy production rate. This gap limits our understanding of entropy production and non-equilibrium effects during droplet coalescence and hinders the broader applicability of existing models in complex systems.

To characterize the rich and complex HNE and TNE effects during droplet merging, and to clarify the driving mechanisms and entropy production processes of droplet coalescence, we resort to the discrete Boltzmann method (DBM)
based on kinetic and mean-field theories. DBM is a non-equilibrium statistical physics modeling and complex physical field analysis method based on the discrete Boltzmann equation\cite{RN705, RN737}. Compared to macroscopic fluid modeling, DBM provides more kinetic information, offering new physical insights for the study of complex systems; and compared to microscopic molecular dynamics, it can simulate and investigate kinetic processes over larger spacetime scales. The core advantage of DBM lies in dynamically and intuitively describing and displaying discrete/TNE effects and behaviors. It addresses two major challenges in numerical experimental research: ``how to model before simulation'' and ``how to analyze after simulation'' for complex flows, especially those with significant discrete state and effects as well as non-equilibrium state and effects\cite{RN737}.

Recent studies have emphasized the importance of investigating TNE effects and behaviors. In scenarios where TNE occurs, neglecting its effects can lead to inaccurate estimations of viscosity, heat flux, and ultimately result in imprecise calculations of macroscopic quantities\cite{RN191, RN27, RN356}. In the field of complex multiphase flows, such as phase separation\cite{RN39, RN40}, bubble/droplet coalescence\cite{RN43, RN198, RN739}, and laser-bubble/droplet interactions\cite{RN730}, TNE effects offer significant advantages in characterizing the evolutionary features of scaled systems. For example, during liquid-vapor phase separation, the maximum points of the total TNE intensity and entropy production rate can be used to distinguish between the spinodal decomposition and domain growth stages\cite{RN39, RN40}. The non-organized momentum flux component in the coalescing direction aids in identifying stages in bubble coalescence\cite{RN198}. Additionally, the intensity of non-organized momentum flux (NOMF) can provide a standard for distinguishing different stages of the droplet collision process and for differentiating between types of droplet collisions\cite{RN43}. In fluid instability systems, by detecting TNE features from different perspectives, a more detailed physical structure of the interface can be provided, which can then be used for the physical identification and tracking technology design of different interfaces within the system\cite{RN32, RN370, RN877, RN878, RN879}.

It is important to note that the droplet coalescence process differs significantly under isothermal and non-isothermal conditions. This difference arises from the distinct coalescence mechanisms and the underlying TNE effects. Understanding the coalescence mechanisms under both isothermal and non-isothermal conditions is crucial for effective droplet manipulation in various scenarios. This paper compares and analyzes the differences and connections between isothermal and non-isothermal droplet coalescence processes, focusing on coalescence mechanisms, morphological features, flow field characteristics, non-equilibrium effects, and entropy production mechanism. The structure of this paper is as follows: Section\ref{Sec2} introduces the DBM method; Section \ref{Sec3} presents simulation results and an analysis of droplet coalescence kinetics under both isothermal and non-isothermal conditions; Section\ref{Sec4} summarizes the findings and outlines future research directions.

\section{The model}\label{Sec2}
The DBM is a kinetic modeling and complex physical field
analysis approach developed based on discrete Boltzmann equation:
\begin{equation}
\frac{{\partial f_{ki}}}{{\partial t}} + \mathbf{v}_{ki} \cdot \frac{{\partial f_{ki}}}{{\partial \mathbf{r}}} = - \frac{1}{\tau} (f_{ki} - f_{ki}^{eq}) + I_{ki},
\end{equation}
where \( f_{ki}^{eq} \) is the discrete form of the local equilibrium distribution  {function \cite{RN361}, with \( k \) representing the \( k \)-th group of particle velocities \( \mathbf{v}_{ki} \), and \( i \) the direction of \( \mathbf{v}_{ki} \)}. The external force term, \( I_{ki} = -[A + \mathbf{B} \cdot (\mathbf{v}_{ki} - \mathbf{u}) + (C + C_q) (\mathbf{v}_{ki} - \mathbf{u})^2] f_{ki}^{eq} \), is introduced using a mean-field approximation \cite{RN191,RN465} to account for the contribution of intermolecular interaction forces to the rate of change of the distribution function. Here, \( A = -2(C + C_q) T \) ensures mass conservation.  \( \mathbf{B} = \frac{1}{\rho T} \bm{\nabla} \cdot \left[ (P - \rho T) \mathbf{I} + \bm{\Lambda} \right] \)with $\bm{\Lambda} = K \bm{\nabla} \rho \bm{\nabla} \rho - K (\rho \bm{\nabla}^2 \rho + \frac{1}{2} |\bm{\nabla} \rho|^2) \mathbf{I} - [\rho T \bm{\nabla} \rho \cdot \bm{\nabla} (K/T)] \mathbf{I}$ representing the density gradient contribution to the pressure tensor, \( \mathbf{I} \) is the unit tensor and \( K \) is the surface tension coefficient. The term $C = \frac{1}{{2\rho {T^2}}}\{ \left( {{P} - \rho
T} \right)\bm {\nabla} \cdot \mathbf{u} + {\bm{\Lambda }}:\bm {\nabla}
{\bf{u}} + a{\rho ^2}\bm {\nabla} \cdot \mathbf{u} - K\left[ {\frac{1}{2}\bm
{\nabla} \rho \cdot \bm {\nabla} \rho \bm {\nabla} \cdot \mathbf{u} + \rho %
\bm {\nabla} \rho \cdot \bm {\nabla} \left( {\bm {\nabla} \cdot \mathbf{u}}
\right) + \bm {\nabla} \rho \cdot \bm {\nabla} {\bf{u}} \cdot \bm {\nabla}
\rho } \right]\}$ represents a partial contribution to energy.
The term \( C_q = \frac{1}{\rho T^2} \bm{\nabla} \cdot (q \rho T \bm{\nabla} T) \) accounts for additional heat conduction within the system. The Carnahan-Starling equation of state (EOS) is given by \( P = \rho T \frac{1 + \eta + \eta^2 - \eta^3}{(1 - \eta)^3} - a \rho^2 \), with \( \eta = \frac{b \rho}{4} \), and \( a \) and \( b \) being the attraction and repulsion parameters, respectively. Here, \( \rho \), \( \mathbf{u} \), and \( T \) represent the local density, velocity, and temperature, respectively.

As a complex physical field analysis method, DBM extracts non-equilibrium information from the system using the high-order non-conserved moment:
\begin{equation}
\begin{aligned}
\bm{\Delta}_{m,n} &= \mathbf{M}_{m,n}(f_{ki} - f_{ki}^{eq}) \\
 = \sum\limits_{ki}& \left( \frac{1}{2} \right)^{1 - \delta_{m,n}}
(f_{ki} - f_{ki}^{eq}) (\mathbf{v}_{ki} \cdot \mathbf{v}_{ki})^{\frac{m-n}{2}} \mathbf{v}_{ki}^{n},
\end{aligned}
\end{equation}
where \( \bm{\Delta}_{m,n} \) describes non-equilibrium effects, including both molecular average behavior and thermal fluctuation behavior—referred to as thermo-hydrodynamic non-equilibrium (THNE) effects. When \( \mathbf{v}_{ki} \) is replaced by \( \mathbf{v}_{ki}^* = \mathbf{v}_{ki} - \mathbf{u} \), \( \bm{\Delta}_{m,n} \) becomes \( \bm{\Delta}_{m,n}^* \), which exclusively represents TNE effects. The difference \( \bm{\Delta}_{m,n} - \bm{\Delta}_{m,n}^* \) represents hydrodynamic non-equilibrium (HNE) effects.

Among the TNE measures, \( \bm{\Delta}_2^* \) represents the NOMF or viscous stress, with its first-order analytical form:
 {$\bm{\Delta}_2^{*(1)} = -\rho T \tau \left[ \bm\nabla \mathbf{u} + (\bm\nabla \mathbf{u})^T - \mathbf{I} \bm \nabla \cdot \mathbf{u} \right]$.}
The term \( \bm{\Delta}_{3,1}^* \) represents the non-organized energy flux (NOEF) or heat flux, with its first-order analytical expression:
$\bm{\Delta}_{3,1}^{*(1)} = -2\rho T \tau \bm{\nabla} T$.
The term \( \bm{\Delta}_3^* \) represents the viscous stress flux, and \( \bm{\Delta}_{4,2}^* \) represents the flux of heat flux. To provide a comprehensive description of the system's non-equilibrium intensity, DBM adopts the non-equilibrium intensity vector:
$\mathbf{S}_{\text{TNE}} = \{ |\bm{\Delta}_2^*|, |\bm{\Delta}_{3,1}^*|, |\bm{\Delta}_3^*|, |\bm{\Delta}_{4,2}^*|, \dots, |\bm{\Delta}_{m,n}^*|, {D^*} \}$,
{with ${D{^{\ast
}=}}\sum_{m,n}|\bm{\Delta}_{m,n}^{\ast }|$ the total TNE intensity.
The definitions, components, and their physical meanings are displayed in the Appendix.}

{Taking the moments of Eq. (1) with the collision invariant vector ($1$, ${\mathbf{v}_{ki}}$, ${{{v}_{ki}^{2}}/2}$), the generalized hydrodynamic equations (GHEs) for non-ideal fluids considering surface tension effect are obtained:
\begin{equation}
{\frac{{\partial \rho }}{{\partial t}}}+\bm {\nabla} \cdot (\rho \mathbf{u})=0,
\end{equation}
\begin{equation}
{\frac{{\partial (\rho \mathbf{u})}}{{\partial t}}}+\bm {\nabla} \cdot (\rho
\mathbf{uu}+P\mathbf{I})+\bm {\nabla} \cdot (\bm{\Lambda }+\bm{\Delta }%
_{2}^{\ast })=0,
\end{equation}
\begin{eqnarray}
\begin{aligned}
{\frac{{\partial {e_{ T}}}}{{\partial t}}}+ \bm {\nabla} \cdot ({e_{T}}\mathbf{u}+P\mathbf{u})\qquad \qquad \qquad \quad \quad \ \quad \, \, \, \qquad \\
+\bm {\nabla} \cdot \left[ {(\bm{\Lambda }+\bm{\Delta }%
_{2}^{\ast })\cdot \mathbf{u}+\bm{\Delta }_{3,1}^{\ast }+2\rho Tq \bm {\nabla} T}\right] =0,
\end{aligned}
\end{eqnarray}
with ${e_{T}}=\rho T-a{\rho ^{2}}+{{K{{\left\vert {\bm {\nabla} \rho }\right\vert }^{2}}}/2+\rho u^{2}/2}$ the total energy density.
Recovering the GHEs represents only one aspect of the physical capabilities of the DBM.
In fact, the full physical description provided by DBM corresponds to the extended hydrodynamic equations \cite{RN191}, which encompass not only the evolution equations for conserved moments, but also for non-conserved moments closely associated with thermodynamic and hydrodynamic non-equilibrium effects.
The significance of these extended equations increases notably as the discrete effects or non-equilibrium intensity becomes stronger.}

 {Both types of non-equilibrium effects, NOMF and NOEF contribute to the entropy production rate,
\begin{equation}
\dot{S}_{\rm pr} = \dot{S}_{\rm NOMF} + \dot{S}_{\rm NOEF},
\end{equation}
where
\begin{equation}
\dot{S}_{\rm NOMF} = \int \left( - \frac{1}{T} \bm{\Delta}_2^* : \bm{\nabla} \mathbf{u} \right) \, d\mathbf{r},
\end{equation}
and
\begin{equation}
\dot{S}_{\rm NOEF} = \int \left( \bm{\Delta}_{3,1}^* + 2 \rho T q \bm{\nabla} T \right) \cdot \bm{\nabla} \left( \frac{1}{T} \right) \, d\mathbf{r}.
\end{equation}
}

\section{Simulations and results}\label{Sec3}

{Droplet coalescence is a multiscale phenomenon driven by the interplay of multiple physical mechanisms, particularly the competing TNE effects. This study systematically explores these mechanisms by examining: (i) slowly varying macroscopic variables, (ii) interface dynamics, (iii) rapidly varying non-equilibrium quantities, and (iv) entropy production rates under both isothermal and non-isothermal conditions.}

All parameters in the manuscript and simulations are dimensionless. The reference variables can be chosen as the reference density $\rho_{\text{r}}$, reference temperature $T_{\text{r}}$, and the global size of the domain $L_{\text{r}}$. The choice of the reference quantity is arbitrary; for instance, the critical density $\rho _{c}$ and the critical temperature $T_c$ can be selected as reference variables, calculated from the non-ideal equation of state. The relationship between the parameters in DBM and the physical ones are as follows:
\begin{equation}
\hat{\rho}=\frac{\rho }{{{\rho _{\text{r}}}}}, \quad \hat{T}=\frac{T}{{{T_{\text{r}}}}}, \quad
\hat{r}_{\alpha }=\frac{r_{\alpha }}{{{L_{\text{r}}}}}, \quad \hat{u}_{\alpha }=\frac{u_{\alpha }}{%
\sqrt{R{T_{\text{r}}}}},
\label{Eq-Dimenless1}
\end{equation}%
\begin{equation}
\left( {\hat{t},\hat{\tau}}\right) =\frac{{\left( {t,\tau }\right) }}{{{%
L_{\text{r}}}/\sqrt{R{T_{\text{r}}}}}},  \quad {\hat{c}_{p}}=\frac{{{c_{p}}}}{R},\quad
\hat{\sigma}=\frac{\sigma }{{{\rho _{\text{r}}}R{T_{\text{r}}}{L_{\text{r}}}}},
\label{Eq-Dimenless2}
\end{equation}%
\begin{equation}
\hat{\mu}=\frac{\mu }{{{\rho _{\text{r}}}{L_{\text{r}}}\sqrt{R{T_{\text{r}}}}}}, \quad
\hat{\kappa}=\frac{\kappa }{{{\rho _{\text{r}}}{L_{\text{r}}}R\sqrt{R{T_{\text{r}}}}}},
\label{Eq-Dimenless3}
\end{equation}%
where $r_{\alpha }$ indicates the space coordinate in the $\alpha$ direction, $c_{p}$ is the specific heat at constant pressure, $\mu$ and $\kappa$ are the viscosity and heat conductivity coefficients, respectively, $\sigma$ is the surface tension, computed as $\sigma =K\int_{-\infty }^{\infty }{{{(\frac{\partial \rho }{\partial {{r}_{\alpha }}})}^{2}}}d{{r}_{\alpha }}$. Dimensionless variables are denoted by a \textquotedblleft $\wedge $\textquotedblright\ sign on the left side.
When the parameters from the non-ideal equation of state (EOS) and the physical property parameters of the real fluid are provided, it is feasible to recover the actual physical quantities from the numerical results \cite{RN40}. In the manuscript, all quantities are dimensionless, and the symbol \textquotedblleft $\wedge $\textquotedblright\ is dropped for simplicity.

To conduct simulations, the particle velocity space is discretized using the D2V33 discrete velocity set \cite{RN361}, while spatial derivatives are discretized using a 16th-order fast Fourier transform (FFT) scheme  \cite{RN83} and temporal derivative is handled using the second-order Runge-Kutta finite difference method. The validity of the physical model and numerical methods has been confirmed in Ref. \onlinecite{RN198}. The computational grid is set to ${N_x} \times {N_y} = 256 \times 256$, with spatial steps of ${\Delta }x = {\Delta }y = {1/128}$. The parameters $a$ and $b$ in EOS are set to $a = 2.0$ and $b = 0.4$, which determine the critical point at ${T_c} = 1.88657$, ${\rho _c}= 1.3044$, and ${P_c} = 0.8832$.

{Two initially stationary droplets of identical size are positioned as:}
\begin{equation}
\begin{array}{l}
\rho (x,y)={\rho _{\rm L}} -\frac{{({\rho _{\rm L}}-{\rho _{\rm V}})}}{2}\tanh \left[ {%
\frac{{\sqrt{{{(x-{x_{0\rm l}})}^{2}}+{{(y-{y_{0\rm l}})}^{2}}}-{r_{0}}}}{{0.5w}}}%
\right] \\
\mathrm{\ \ \ \ \ \ \ \ \ \ \ \ \ \ \ }-\frac{{({\rho _{\rm L}}-{\rho _{\rm V}})}}{2}\tanh %
\left[ {\frac{{\sqrt{{{(x-{x_{0\rm r}})}^{2}}+{{(y-{y_{0\rm r}})}^{2}}}-{r_{0}}}}{{%
0.5w}}}\right] .
\end{array}
\end{equation}
Here, \(\rho_{\rm L} = 2.0658\) and \(\rho_{\rm V} = 0.6894\) are the densities of the liquid and vapor phases at a temperature of \(T = 1.8\). The total width of the boundary layer is \(2w = 12\Delta x\). The initial radius of both droplets is the same, denoted as \(r_0\). The center coordinates of the left and right droplets are \((x_{0\rm l}, y_{0\rm l})\) and \((x_{0\rm r}, y_{0\rm r})\), respectively. The distance between the two centers is \(\left| x_{0\rm l} - x_{0\rm r} \right| = 2r_0 + r_d\). For all cases, \(y_{0\rm l} = y_{0\rm r}\) and \(r_d = \omega \Delta x\), where \(\omega\) is a real constant. The relaxation time used in the simulation is \(\tau = 0.0025\), with \(\Pr = 0.5\). Unless otherwise specified, \(K = 0.0001\).

\subsection{Differences in HNE effects and coalescence mechanisms}

\begin{figure*}[htbp]
\centering\includegraphics*
[width=0.8\textwidth,trim=0.1 0.1 0.1 0.1,clip]{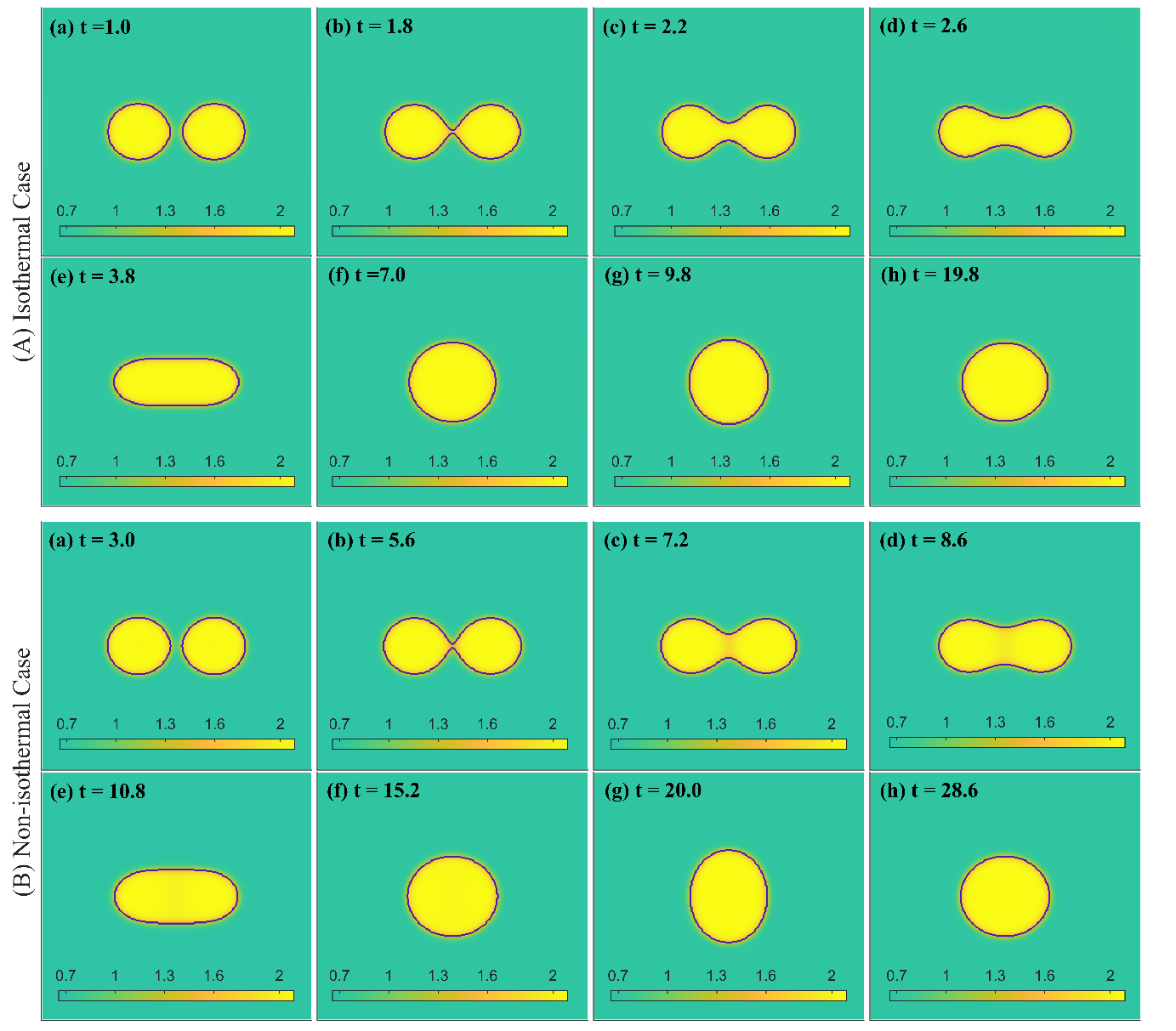}
 \caption{\centering{{Density distribution at eight characteristic moments with similar morphology under isothermal and non-isothermal conditions:} (A) Isothermal case; (B) Non-isothermal case. Here, each sub-figure represents a partial simulated area with grid dimensions ${N_x} \times {N_y} = 256 \times 256$.}}
 \label{F1}
\end{figure*}
Figure \ref{F1} shows the spatial distribution of density at eight characteristic moments with similar morphology under isothermal and non-isothermal conditions. The droplet coalescence process is significantly slower in the non-isothermal case compared to the isothermal case. Specifically, {under non-isothermal condition}, the initiation of the coalescence process is delayed, cut-through occurs more slowly, the rapid coalescence phase lasts longer, and exhibits noticeable damped oscillations.
 \begin{figure*}[htbp]
\centering\includegraphics*
[width=0.9\textwidth,trim=0.1 0.1 0.1 0.1,clip]{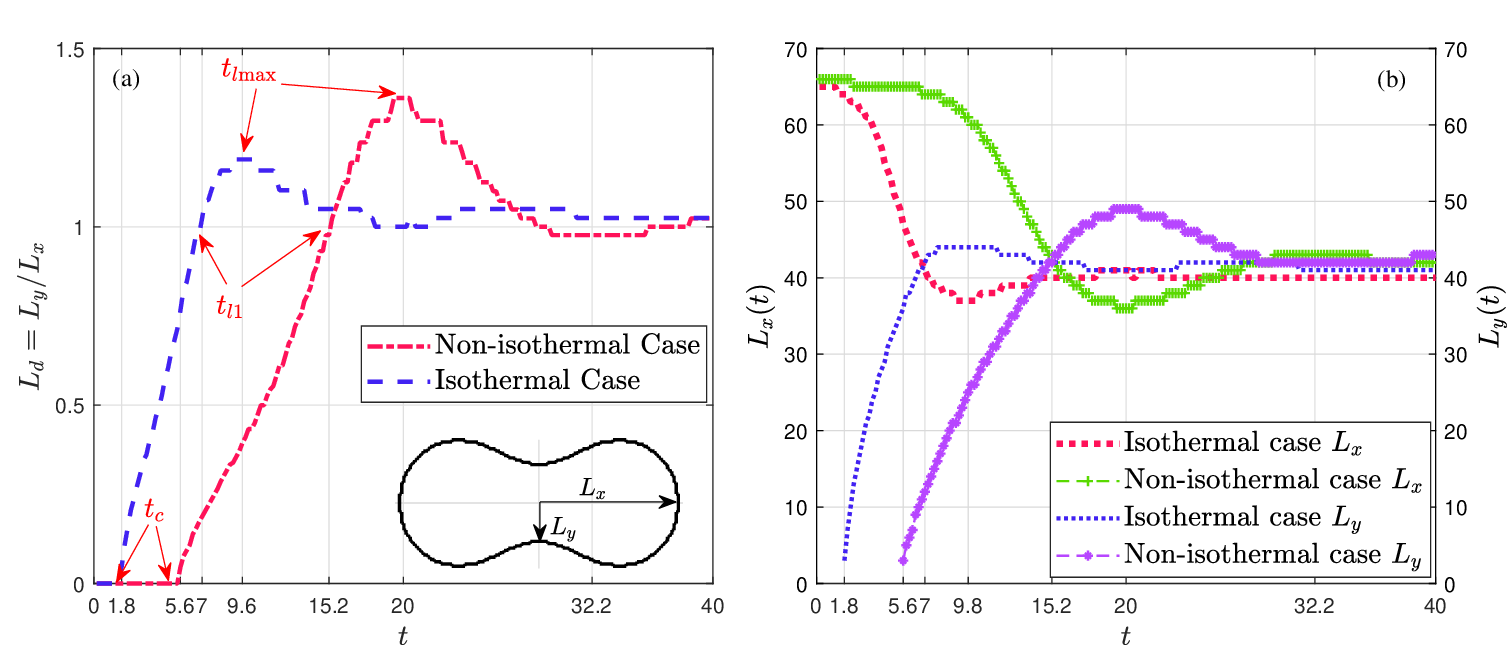}
 \caption{\centering{Time evolution of the morphological factor $L_d = L_y/L_x$ (a), the long semi-axis $L_x$ and the short semi-axis $L_y$ (b) under isothermal and non-isothermal conditions. The slope of \( L_d \) can be used to characterize the droplet coalescence speed.}}
 \label{F2}
\end{figure*}
\begin{figure*}[htbp]
\centering\includegraphics*
[width=0.8\textwidth,trim=0.1 0.1 0.1 0.1,clip]{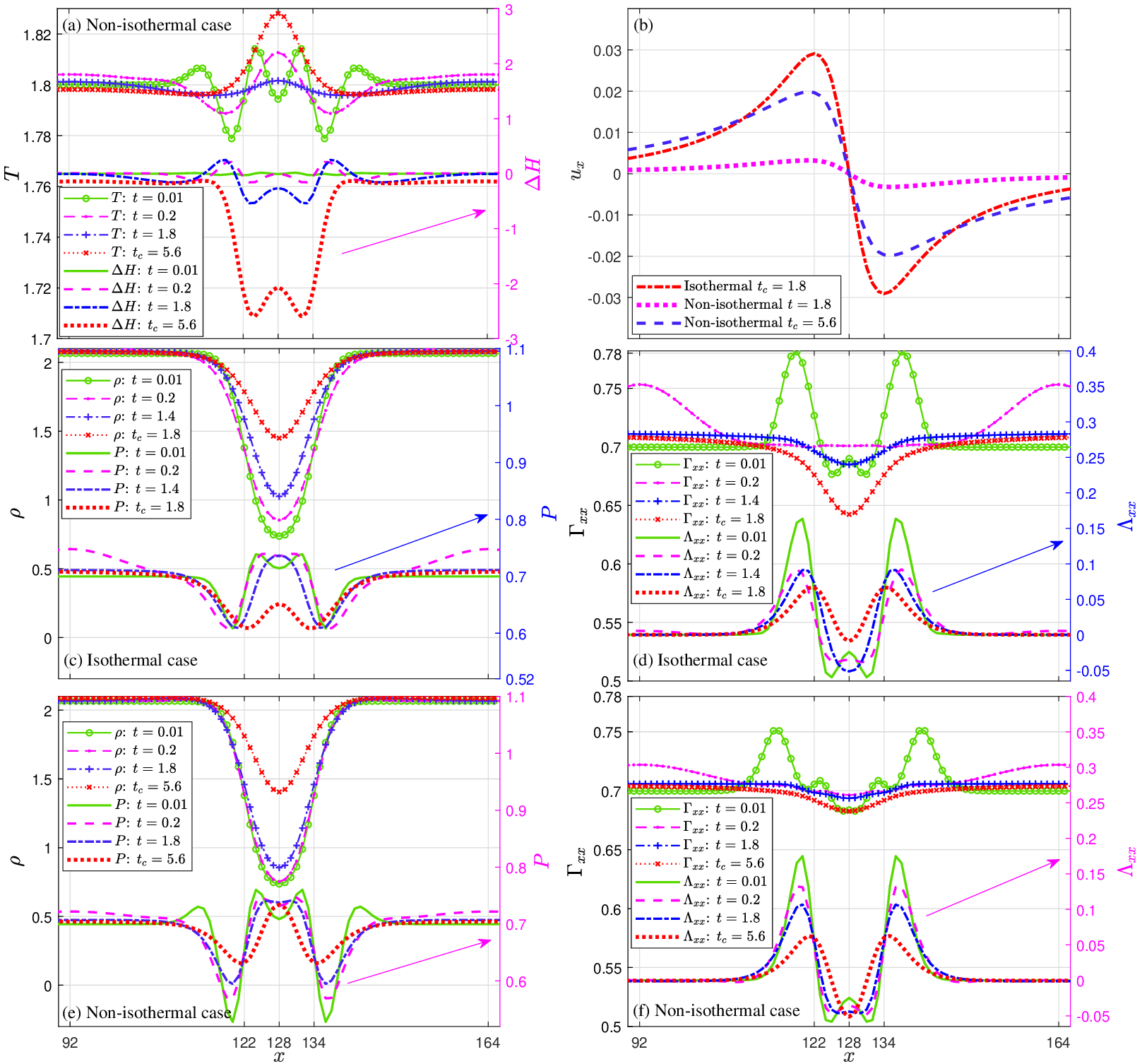}
 \caption{\centering{(a) Profiles of $T$ and  $\Delta H$ under non-isothermal condition; (b) Profiles of $u_x$ under both isothermal and non-isothermal conditions; (c)-(f) Profiles of $\rho$, $P$, $\Gamma_{xx}$, and  ${{\Lambda }_{xx}}$ under both isothermal and non-isothermal conditions. In a phase-separating system, the gradient of the total pressure tensor $\bm{\nabla} \bm{\Gamma }$ serves as the overall driving force for droplet coalescence, the gradient of $\bm{\nabla} \bm \Lambda$ (i.e., surface tension) provides the driving force, while the gradient of pressure $\bm{\nabla} P$ acts as the resistance.}}
 \label{F3}
\end{figure*}

\textbf{(i) Slower initiation and cut-through processes: }
The coalescence initiation duration {under non-isothermal condition is much longer than that under isothermal condition}. For the isothermal case, the cut-through time of two static droplets is \( t_c = 1.8 \) [see Fig. \ref{F1} Ab], while for the non-isothermal case, it is \( t_c = 5.6 \) [see Fig. \ref{F1} Bb]. The time difference in both cases is \( \Delta t_c = 3.8 \).
The differences in cut-through time can be more obviously reflected in the evolution of morphological factor. As shown in Fig. \ref{F2}, in both cases, at \( t = t_c \), the morphological factor \( L_d = L_y/L_x \) and \( L_y \) begin to increase, where \( L_x \) and \( L_y \) are the long and short semi-axes of the new droplet, respectively.

\textbf{(ii) Longer duration of the rapid coalescence process}: As shown in Fig. \ref{F1} and Fig. \ref{F2}, {under isothermal condition}, the time at which the new droplet first forms an unstable spherical shape, corresponding to the morphological factor \( L_d \) first equaling 1, is \( t_{l1} = 7.0 \). Then the time difference between cut-through and the first formation of an unstable spherical droplet is \( \Delta t_{l1c} = t_{l1} - t_c = 5.2 \).
Under non-isothermal condition, the time for the new droplet to first forms an unstable spherical shape is \( t_{l1} = 15.2 \), and the time difference between cut-through and the first formation of an unstable spherical droplet is \( \Delta t_{l1c} = t_{l1} - t_c = 9.6 \). The time difference for the first formation of an unstable spherical droplet in both cases is { \( \Delta t_{l1} = 8.2 \). Under non-isothermal condition}, the growth rate of the morphological factor \( L_d \) of the new droplet is smaller than {under isothermal condition}, because the slopes of both the long semi-axis \( L_x \) and the short semi-axis \( L_y \) of the droplet are smaller { under non-isothermal condition} [see Fig. \ref{F2}].

\textbf{(iii) Greater oscillation amplitude and longer oscillation period}:
 {After coalescence, the newly formed larger droplet often exhibits a highly non-circular shape. Since the system is not initially at its minimum surface energy, surface tension drives the droplet toward restoring a circular configuration.
During this relaxation process, due to fluid inertia, the droplet typically does not reach equilibrium in a single step. Instead, it undergoes damped oscillations around the spherical shape until its kinetic energy is fully dissipated.
The timescale of these oscillations is governed jointly by viscous dissipation, heat transfer, and surface tension effects \cite{RN320}.}
As shown in Fig. \ref{F2}(a),  {under isothermal condition}, the time at which \( L_d \) reaches its maximum is \( t_{l\rm max} = 9.6 \), whereas  {under non-isothermal condition}, \( t_{l\rm max} = 20.0 \), with a difference of \( \Delta t_{ l\rm max } = 10.4 \). After the first oscillation period, under isothermal condition, the oscillation amplitude of \( L_x \) and \( L_y \) is almost zero, as shown in Fig. \ref{F2}(b). In contrast,  {under non-isothermal condition}, small oscillations persist. Furthermore, the maximum value of \( L_d \)  {under non-isothermal condition} is greater than that under isothermal condition. This can be attributed to the more complex stress and flow field distributions near the vapor-liquid interface  {under non-isothermal condition}, which prevent the liquid-vapor interface from stabilizing easily \cite{RN780}.
\begin{figure*}[htbp]
\centering\includegraphics*
[width=0.8\textwidth,trim=0.1 0.1 0.1 0.1,clip]{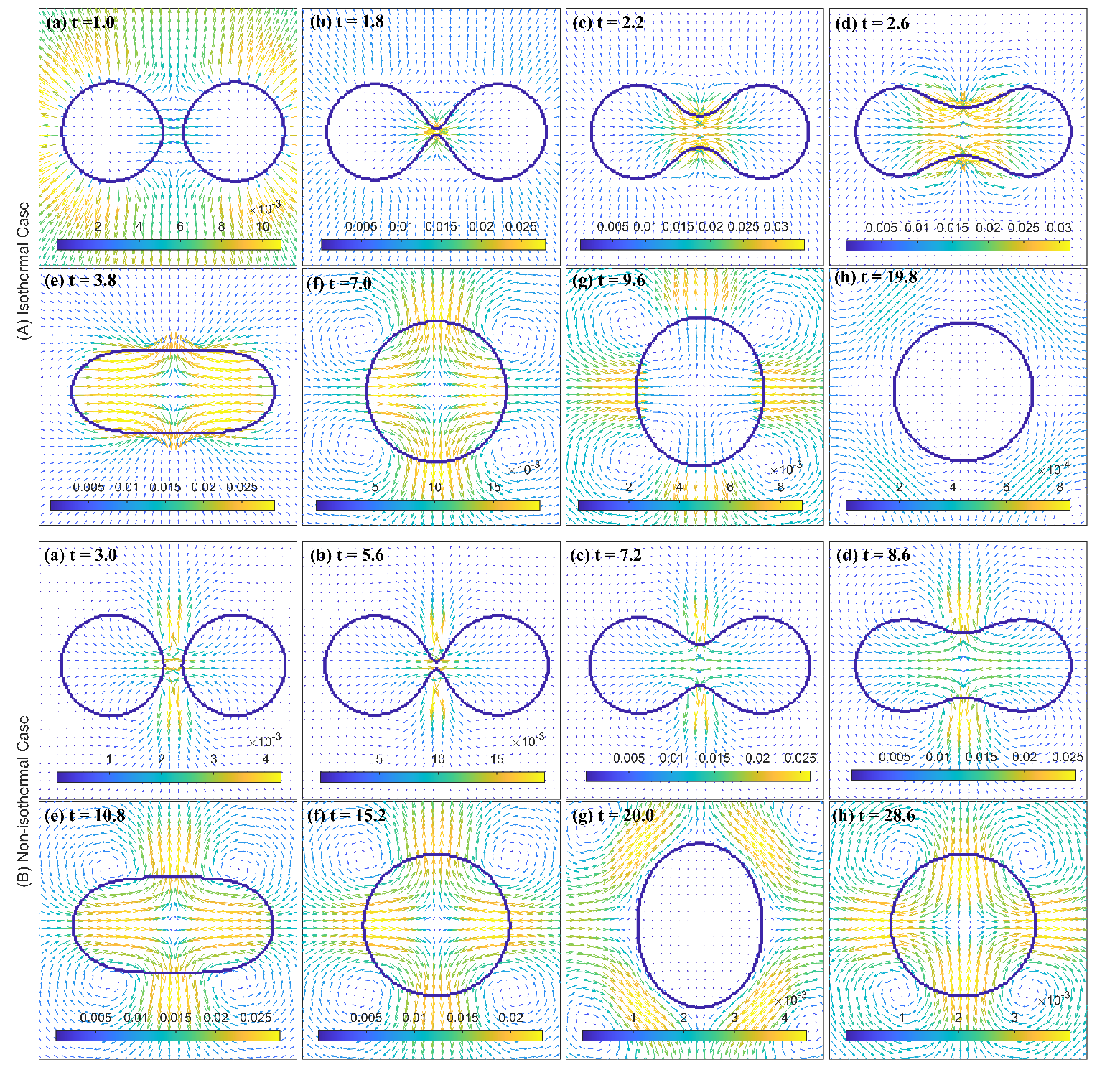}
    \caption{Spatial distribution of the flow field at eight characteristic moments with similar morphology under isothermal and non-isothermal conditions: (A) Isothermal case; (B) Non-isothermal case. The blue solid lines represent the liquid-vapor interface. Here, each sub-figure represents a partial simulated area with ${N'_x} \times {N'_y} = 150 \times 150$.}
    \label{F4}
\end{figure*}

To explain the above differences, Fig. \ref{F3} presents the spatial distribution along the horizontal midline of various quantities: temperature \(T\), latent heat of phase change \(\Delta H = H_t - H_0\)  {(with enthalpy \(H = \epsilon + P/\rho\), the internal energy density ${\epsilon}=\rho T-a{\rho ^{2}}+{{K{{\left\vert {\bm {\nabla} \rho }\right\vert }^{2}}}/2}$, and $H_t$ and $H_0$ being the enthalpies at time $t$ and initial state)}, velocity in the coalescence direction \( u_x \), density \(\rho\), pressure \(P\), the \(xx\)-component of the total pressure tensor \( \bm{\Gamma} = P \mathbf{I} + \bm{\Lambda} \) (denoted as \( \Gamma_{xx} \)), and the \(xx\)-component of the pressure tensor contribution from the macroscopic gradient \(\bm{\Lambda}\) (denoted as \( \Lambda_{xx} \)).

In the isothermal case, it is assumed that the system is in contact with a sufficiently large heat bath, resulting in instantaneous heat exchange with the environment \cite{RN83}. This assumption leads to no temperature gradient within the system. In contrast, the non-isothermal case incorporates the temperature field's inhomogeneity, introducing additional HNE and TNE effects, such as the temperature-induced Marangoni effect and heat conduction. These effects significantly alter the system's dynamics and flow field morphology.

As shown in Fig. \ref{F3}(a), the release of latent heat in the non-isothermal case creates a localized high-temperature region in the liquid bridge between the droplets, leading to a prolonged local increase in pressure \(P\). In the isothermal case, the pressure peak at the center of the liquid bridge oscillates but gradually decreases after \(t \geq 0.2\), reducing by approximately $11.5\%$ at \(t = 1.8\). In contrast, the pressure peak in the non-isothermal case remains nearly constant during the period \(0.2 \leq t \leq t_c = 5.6\).

The pressure gradient in the non-isothermal case is larger, resulting in stronger resistance to droplet coalescence. Although the gradient of surface tension \( \bm{\nabla} \bm{\Lambda} \) is larger in the non-isothermal case (with larger positive peaks and smaller negative troughs, as shown by the curves on the right axis of Fig. \ref{F3}(d) and (f)), the gradient of the total pressure tensor \( \bm{\nabla} \bm{\Gamma} \) is greater in the isothermal case [as shown by the curves on the left axis of Fig. \ref{F3}(b) and (d)]. This explains why the coalescence initiation time is significantly prolonged in the non-isothermal case.

As shown in Fig. \ref{F3}(b), at the same moment and state, the flow velocities inside and outside the liquid-vapor boundary layer in the non-isothermal case are consistently much smaller than those in the isothermal case. Even at \(t = t_c = 5.6\), the coalescence velocity is still smaller compared to \(t = t_c = 1.8\) in the isothermal case.

Interestingly, for \( t \leq t_c \), while the pressure gradient decreases in both cases, the manner in which it decreases is distinctly different. In the isothermal case, the pressure gradient decreases by maintaining the trough value approximately constant, while the pressure peak in the liquid bridge region gradually decreases. In contrast, in the non-isothermal case, the pressure peak remains high, and the trough value gradually increases. This behavior indicates that, in the droplet coalescence process, the pressure gradient influences the expansion of the liquid-vapor interface in two opposite ways in the isothermal and non-isothermal cases. This distinction is more clearly observed in the velocity field, as shown in Fig. \ref{F4}.

\begin{figure}[htbp]
\centering\includegraphics*
[width=0.5\textwidth,trim=0.1 0.1 0.1 0.1,clip]{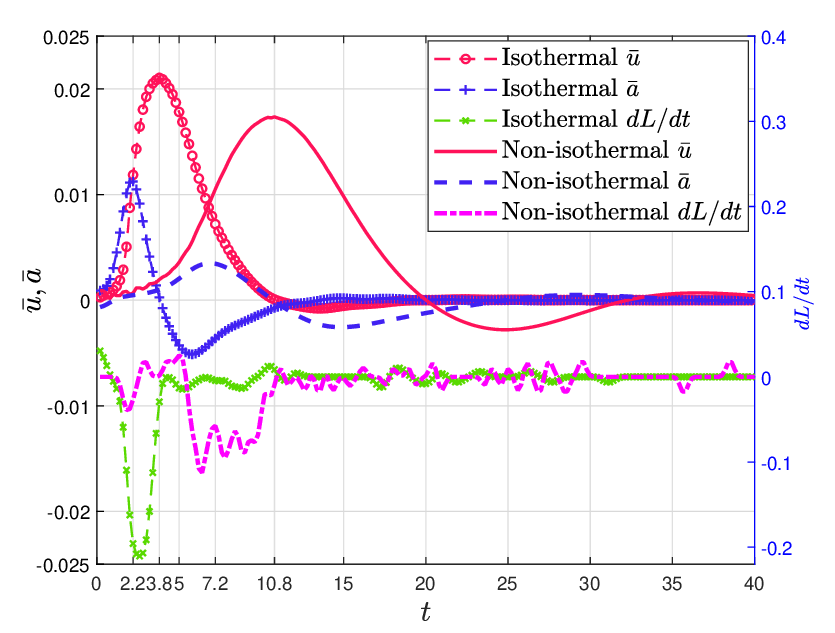}
 \caption{\centering{Time evolution of the average coalescence velocity $\bar u$, average coalescence acceleration $\bar a$ and $dL/dt$, under both isothermal and non-isothermal conditions. Here, $\bar{u}$ is the statistical average of the velocity  {in the coalescencing direction within the left half region} $\bar{u}={\sum{\rho (x,y,t){{u}_{x}}(x,y,t)}}/{\sum{\rho (x,y,t)}}$ and $\bar{a}=d\bar{u}/dt$.}}
 \label{F5}
\end{figure}

Figure \ref{F4} illustrates the spatial distribution of the flow field at eight characteristic moments with similar morphology, under both isothermal and non-isothermal conditions. The blue solid lines represent the liquid-vapor interface during coalescing. A comparative analysis of the flow field distribution in Fig. \ref{F4} reveals significant differences in the velocity vortices between the two conditions.

(i) Under isothermal condition, when \( t \leq 3.8 \), the velocity in the vapor phase along the vertical direction of coalescence (the \(y\)-direction) converges toward the saddle point in the saddle region between the two droplets. No vortices are formed at the liquid-vapor interface in this region. In contrast,  {under non-isothermal condition}, when \( t \leq 10.8 \), the velocity in the vapor phase outside the saddle point diverges along the vertical direction of coalescence. This divergence creates vortices with opposite rotational directions at the liquid-vapor interface in the saddle region. The non-isothermal case results in a complex flow pattern near the interface, with the Marangoni effect significantly enhancing the flow outside the interface.

(ii) Under isothermal condition, due to the faster release of surface energy, the peak of the velocity in merging direction is higher compared to the non-isothermal case. As shown in Fig. \ref{F5}, the coalescence speed  {under isothermal condition} starts faster and reaches its peak earlier. Specifically, the time when the coalescence speed reaches its maximum for the isothermal case is $t_{u\rm max} = 3.8 $, while for the non-isothermal case,  $ t_{u \rm max} = 10.8  $. Furthermore, the peak value  $ \bar{u}_{\rm max}  $ is larger in the isothermal case. However,  {under isothermal condition}, the coalescence speed decays more rapidly, particularly during the damping oscillation phase ( $ t \geq t_{l1}  $). After the first oscillation cycle, the coalescence speed  {under isothermal condition} almost decays to zero, while  {under non-isothermal condition}, the coalescence speed continues to oscillate for several cycles before gradually decaying to zero.

\subsection{Differences in TNE effects}

The upper section discusses the differences in HNE between isothermal and non-isothermal cases during droplet coalescence. Below, we analyze the more fundamental TNE differences, as TNE is both the source of HNE.
\begin{figure*}[htbp]
\centering\includegraphics*
[width=1.\textwidth,trim=0.1 0.1 0.1 0.1,clip]{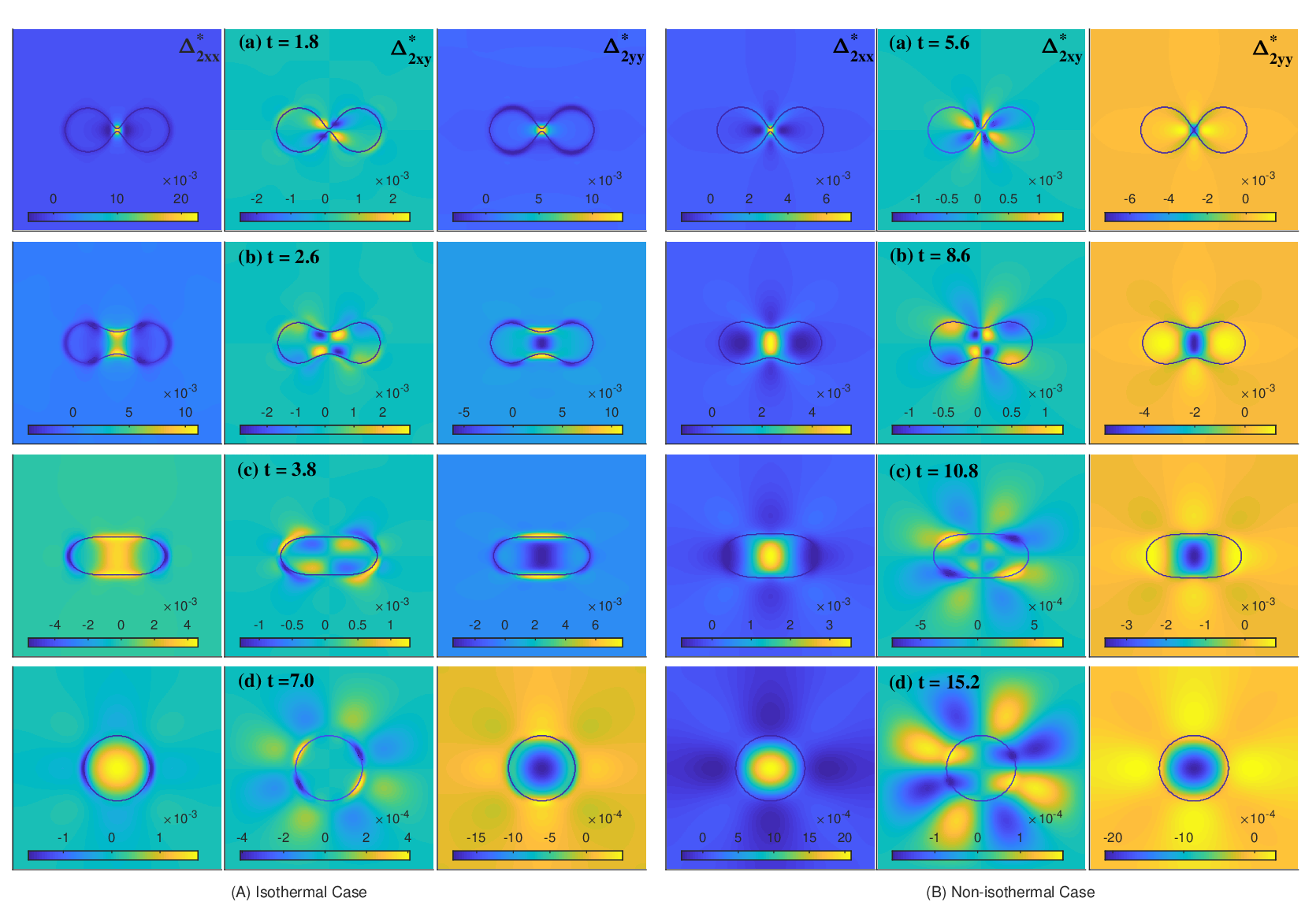}
 \caption{\centering{Spatiotemporal evolution characteristics of the system's NOMF: (A) Isothermal condition, (B) Non-isothermal condition. Each sub-figure represents a partial simulated area with ${N_x} \times {N_y} = 256 \times 256$.}}
 \label{F6}
\end{figure*}
\begin{figure*}[htbp]
\centering\includegraphics*
[width=1.\textwidth,trim=0.1 0.1 0.1 0.1,clip]{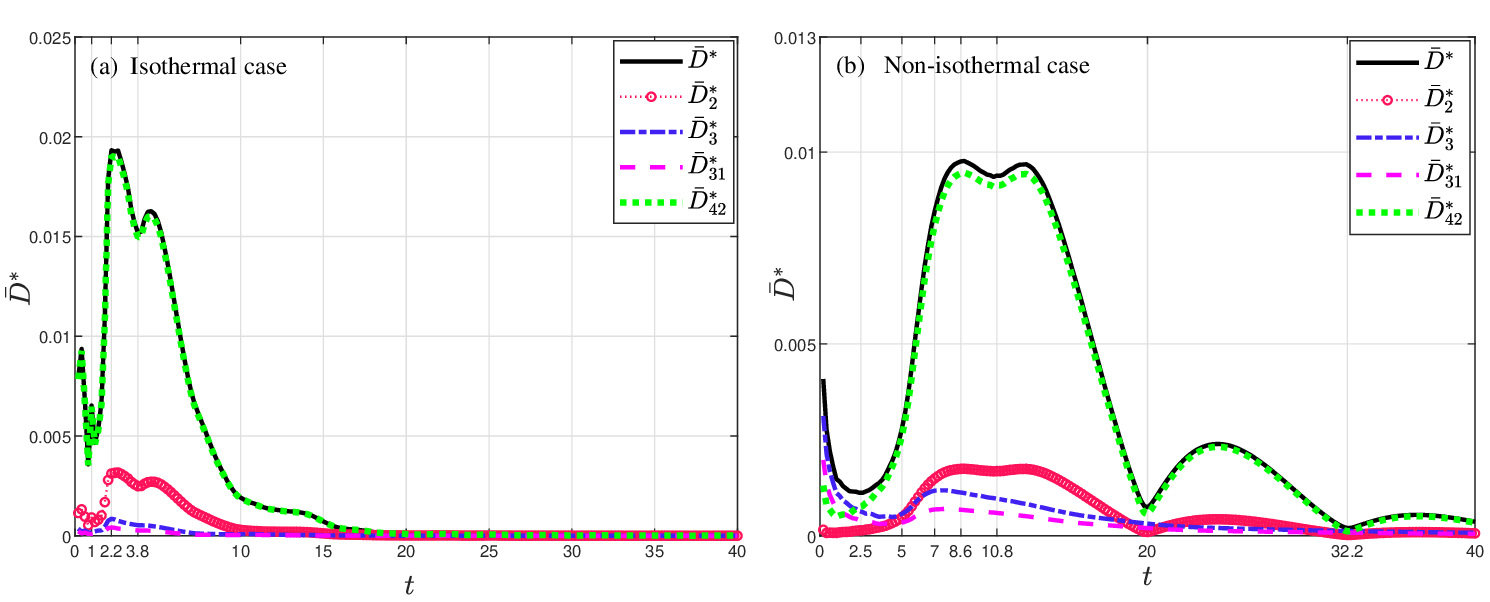}
 \caption{\centering{Time evolution of the total non-equilibrium strength $\bar D^{\ast}$, NOMF strength $\bar D^{\ast}_{2}$, NOEF strength $\bar D^{\ast}_{3,1}$, strength of flux of NOMF $\bar D^{\ast}_{3}$, and strength of flux of NOEF $\bar D^{\ast}_{4,2}$: (a) Isothermal case; (b) Non-isothermal case.}}
 \label{F7}
\end{figure*}

Figure \ref{F6} shows the spatial distribution of the three components of the NOMF, $\bm{\Delta^{\ast}}_2$, at four characteristic moments under both isothermal and non-isothermal conditions. As a TNE characteristic, the spatiotemporal evolution of the flow field in both cases governs the distribution and evolution of $\bm{\Delta^{\ast}}_2$, since $\bm{\Delta^{\ast}}_2$ primarily driven by the velocity gradient. As a TNE effect, $\bm{\Delta^{\ast}}_2$ influences the flow field evolution at faster time scales and smaller spatial scales.

The key differences between the two cases are as follows: (i) In the isothermal case, the spatial distributions of $\Delta^{\ast}_{2xx}$ and $\Delta^{\ast}_{2yy}$ are not completely opposite due to the different directions of the velocity vortices. In contrast, these distributions are fully antisymmetric in the non-isothermal case. (ii) In the non-isothermal case, the spatial distribution of $\Delta^{\ast}_{2xy}$ is more complex due to the increased variety of non-equilibrium driving forces. $\Delta^{\ast}_{2xy}$ in the non-isothermal case evolves from an interior-exterior antisymmetric double quadrupole, while in the isothermal case, it evolves from an antisymmetric single quadrupole structure. The coupling of multiple TNE effects in the non-isothermal system makes the spatiotemporal evolution of non-equilibrium behaviors more complex.

Figure \ref{F7} illustrates the time evolution of several non-equilibrium intensities, including the total non-equilibrium intensity $\bar{D}^{\ast}$, NOMF intensity $\bar{D}^{\ast}_{2}$, NOEF intensity $\bar{D}^{\ast}_{3,1}$, strength of flux of NOMF $\bar D^{\ast}_{3}$, and strength of flux of NOEF $\bar D^{\ast}_{4,2}$ in both isothermal and non-isothermal systems.
The main differences between the two cases are as follows:

\textbf{(i) Duration and peak of TNE effects:}
In the isothermal case, non-equilibrium effects last for a shorter time but have higher peak values. For example, all intensities reach their maximum at approximately $t = 2.2$ for the isothermal case, which is earlier than in the non-isothermal case, where they peak at $t = 8.6$. Additionally, the peak value of $\bar{D}^{\ast}$ is about twice as large in the isothermal case compared to the non-isothermal case.
Similar to the non-isothermal case \cite{RN739}, in the isothermal condition, the total TNE intensity $\bar D^{\ast}$ reaches its maximum value at $t = 2.2$, when the boundary length decays the fastest and the merging acceleration is the greatest; it reaches its first local trough at $t = 3.8$, when the merging velocity is the highest. That is, the characteristic points on the evolution curve of $\bar D^{\ast}$ can be used to delineate the various stages of the droplet coalescence process.

\textbf{(ii) Damping oscillation:}
The non-isothermal case exhibits more pronounced damping oscillations in $\bar{D}^{\ast}_{2}$, $\bar{D}^{\ast}_{4,2}$, and $\bar{D}^{\ast}$ compared to the isothermal case, with larger amplitudes, longer periods, and extended durations.

\textbf{(iii) Interplay of various TNE effects:}
In the isothermal case, there is no temperature gradient, so $\bar{D}^{\ast}_{3}$ and $\bar{D}^{\ast}_{3,1}$ tend to zero, and the total non-equilibrium intensity is mainly driven by the velocity gradients ($\bar{D}^{\ast}_{2}$ and $\bar{D}^{\ast}_{4,2}$).
In the non-isothermal case, when $t \leq 5.0$, the total non-equilibrium intensity is dominated by $\bar{D}^{\ast}_{3}$ and $\bar{D}^{\ast}_{3,1}$ due to the temperature gradient. After $t = 5.0$, the velocity gradients increase, and $\bar{D}^{\ast}_{2}$ and $\bar{D}^{\ast}_{4,2}$ take over as the dominant contributors. \textbf{\emph{The alternation between the dominance of non-equilibrium behaviors driven by temperature gradients  and velocity gradients  corresponds to the moment when the two droplets begin to make contact.}}

This is because, in the early stage of the relaxation process, when the droplets approach each other, the velocity gradient is negligible, and the temperature gradient due to latent heat of phase change governs the system's evolution. After the droplets make contact, the relative velocity between particles increases, causing the velocity gradient to dominate. As shown in Fig. \ref{F5}, at $t = 5.0$, the droplets begin to make contact but have not yet merged, marking the moment when the boundary length rapidly decreases.  {In the later stage} of coalescence, $\bar{D}^{\ast}_{3,1}$ and $\bar{D}^{\ast}_{3}$ decay smoothly to zero, while $\bar{D}^{\ast}_{2}$ and $\bar{D}^{\ast}_{4,2}$ exhibit damped oscillations and also decay to zero as the droplet shape evolves.
In summary, the increasing number of driving factors in the non-isothermal case adds the interplay and complexity to the system, as evidenced by the alternating dominance of behaviors driven by temperature gradients ($\bm{\Delta}^{\ast}_{3}$, $\bm{\Delta}^{\ast}_{3,1}$) and velocity gradients ($\bm{\Delta}^{\ast}_{2}$, $\bm{\Delta}^{\ast}_{4,2}$).

 {\textbf{(v) Higher-order TNE effects in the isothermal case:}
Theoretically, the non-organized energy flux (NOEF) contains contributions from various orders of non-equilibrium effects, as expressed by
$\bm{\Delta}_{3,1}^* = \bm{\Delta}_{3,1}^{*(1)} + \bm{\Delta}_{3,1}^{*(2)} + \cdots$,
where \( \bm{\Delta}_{3,1}^* = \bm{\Delta}_{3,1}^{*(1)} \) in the first-order model. According to theoretical derivation, \( \bm{\Delta}_{3,1}^{*(1)} \) only includes contributions from the temperature gradient, whereas \( \bm{\Delta}_{3,1}^{*(2)} \) incorporates both temperature and velocity gradient contributions \cite{RN191}.
Thus, in the isothermal case, despite the absence of a temperature gradient, \( \bar{D}^{\ast}_{3,1} \) and \( \bar{D}^{\ast}_{3} \) are not exactly zero. This phenomenon reflects the complex cross-coupling of multiple physical quantities in constitutive relations, which is one manifestation of the multiscale system's complexity.
This also indicates that the current DBM framework primarily focuses on first-order non-equilibrium effects, but naturally captures partial second-order non-equilibrium behaviors without requiring additional modifications.}

 {Overall, as a phenomenon, non-equilibrium fundamentally embodies system complexity by revealing the nonlinear coupling among various physical mechanisms. As a driving effect, non-equilibrium significantly influences system evolution through its intrinsic, dominant, and integrative characteristics. Non-equilibrium quantities thus serve as sensitive probes for identifying competing mechanisms, providing earlier, more sensitive, and more precise detection of dominant processes and their transitions compared to conventional macroscopic variables.}

\subsection{Differences in entropy production mechanisms}

\begin{figure*}[htbp]
\centering\includegraphics*
[width=1.\textwidth,trim=0.1 0.1 0.1 0.1,clip]{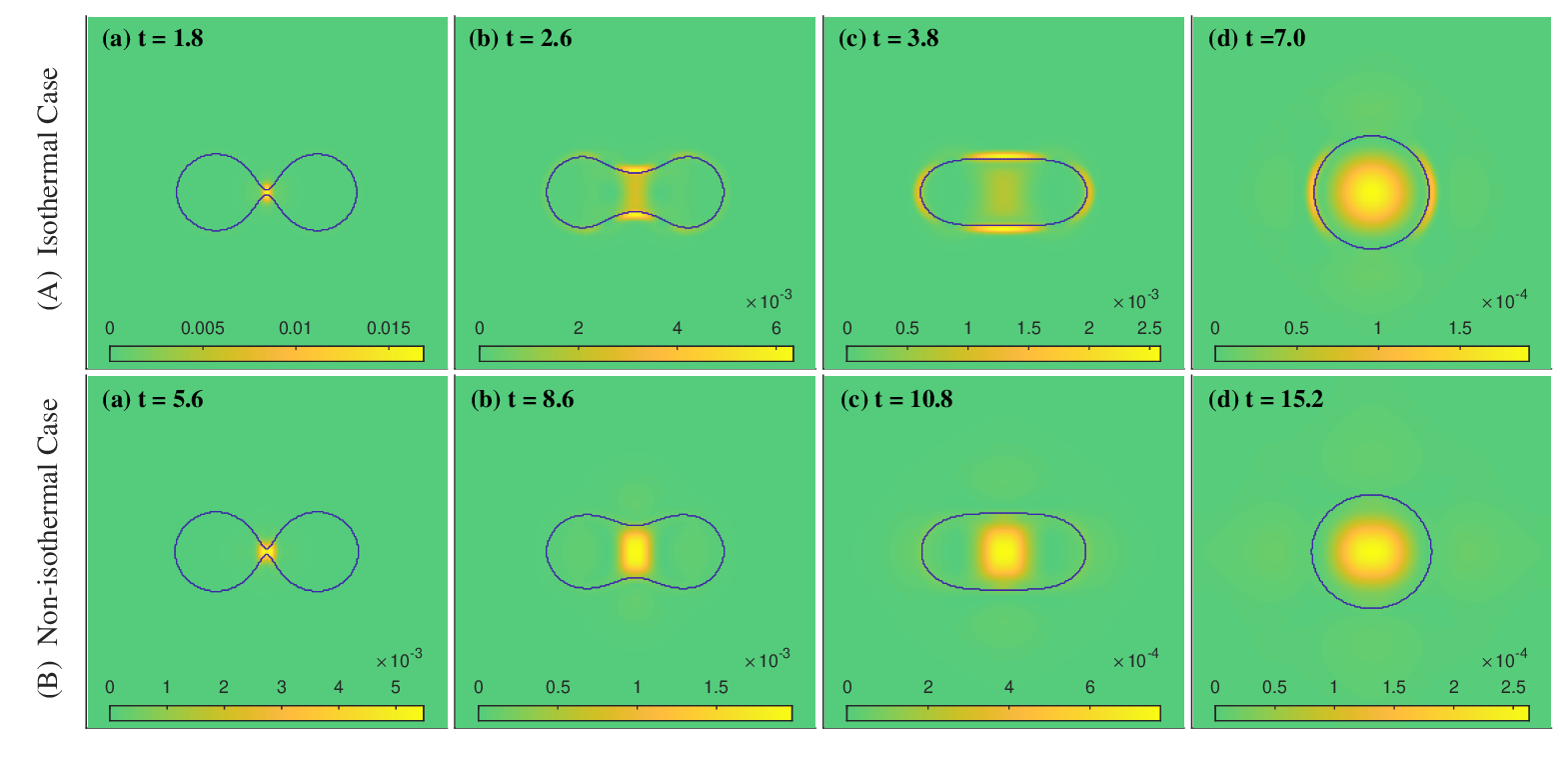}
 \caption{\centering{Spatial distribution of the total entropy production rate $\bar {\dot S}_{\rm pr}$ at four characteristic moments with similar morphology under isothermal and non-isothermal conditions: (A) Isothermal case; (B) Non-isothermal case. Here, each sub-figure represents a partial simulated area with ${N_x} \times {N_y} = 256 \times 256$.}}
 \label{F8}
\end{figure*}
Figure \ref{F8} illustrates the spatial distribution of the total entropy production rate ${{\dot{S}}_{\text{pr}}}$ at four characteristic times, showing similar morphology for both isothermal and non-isothermal conditions. For $t \le t_c$, ${{\dot{S}}_{\text{pr}}}$ is mainly concentrated in the quasi-contact region between the two droplets in both two cases, as shown in Fig. \ref{F8}.
The difference lies in the fact that ${{\dot{S}}_{\text{pr}}}$ in the isothermal case is more concentrated, with its peak about three times higher than in the non-isothermal case. During the rapid coalescence stage ($t_c \le t \le t_{l1}$), ${{\dot{S}}_{\text{pr}}}$ in the isothermal case is primarily concentrated on the outer side of the saddle region of the liquid-vapor boundary layer between the two droplets. In contrast, in the non-isothermal case, ${{\dot{S}}_{\text{pr}}}$ is mainly concentrated in the liquid bridge region between the two droplets.

The reasons are as follows. In the isothermal case, ${{\dot{S}}_{\text{pr}}}$ is solely influenced by NOMF where the velocity gradient is concentrated corresponds to the main distribution of entropy production. In the non-isothermal case, ${{\dot{S}}_{\text{pr}}}$ is influenced by both NOMF and NOEF.
The inhomogeneous temperature field caused by latent heat broadens the entropy production region,
 while simultaneously suppressing local flow velocity, viscous stress, and the resulting entropy production.
\begin{figure}[htbp]
\centering\includegraphics*
[width=0.5\textwidth,trim=0.1 0.1 0.1 0.1,clip]{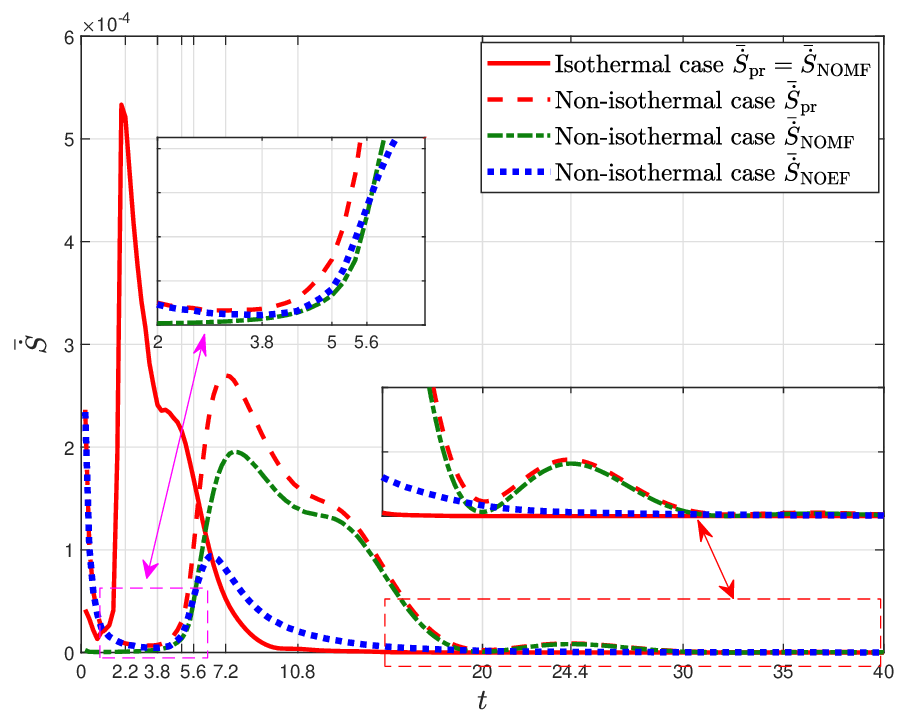}
 \caption{\centering{ {Temporal evolution of the entropy production rate under isothermal and non-isothermal conditions.}}}
 \label{F9}
\end{figure}
\begin{figure}[htbp]
\centering\includegraphics*
[width=0.5\textwidth,trim=0.1 0.1 0.1 0.1,clip]{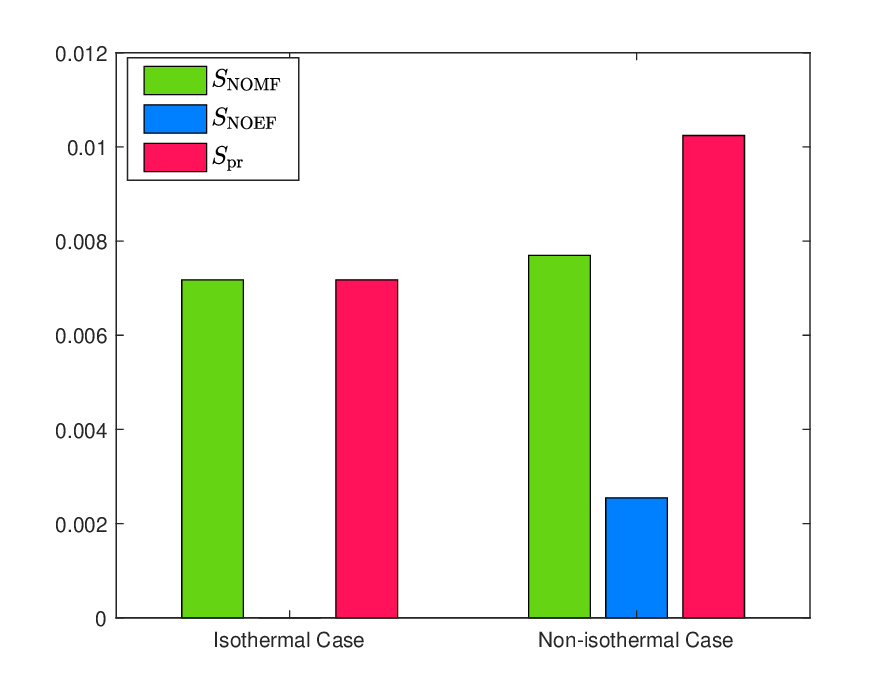}
 \caption{\centering{Total entropy production during coalescence under both isothermal and non-isothermal conditions.}}
 \label{F10}
\end{figure}

Figure \ref{F9} presents the entropy production rate evolution curves over time for both isothermal and non-isothermal conditions. A detailed comparison of the total entropy production rate $\bar{\dot{S}}_{\text{pr}}$ shows distinct behaviors between the two scenarios. In the isothermal case, $\bar{\dot{S}}_{\text{pr}}$ reaches its peak at $t = 2.2$, with a higher peak value compared to the non-isothermal case. However, it decays rapidly, displaying a relatively smooth, unimodal structure with small fluctuations and stabilizing after $t \geq 10$.
In contrast, the non-isothermal case reveals a more intricate evolution pattern. Although the initial peak is lower than in the isothermal case, the system experiences significant fluctuations around $t = 7.2$ and $t = 24.4$, leading to a multi-peak structure. This behavior suggests that, in the non-isothermal case, the disturbances induced by the temperature gradient alongside the velocity gradient contribute to extended and prolonged non-equilibrium dissipation processes.
In the isothermal case, the isothermal entropy production curve can also be used to identify various stages of droplet coalescence: the peak point ($t = 2.2$) and the first local trough point ($t = 3.8$) on the evolution curve correspond to the moments of maximum merging acceleration and maximum merging velocity, respectively.

Comparing the entropy production rates of the NOMF, ${{\bar{\dot{S}}}_{\text{NOMF}}}={\sum{\rho {{{\dot{S}}}_{\text{NOMF}}}}}/{\sum{\rho }}$, and that of  the NOEF, ${{\bar{\dot{S}}}_{\text{NOEF}}}={\sum{\rho {{{\dot{S}}}_{\text{NOEF}}}}}/{\sum{\rho }}$, in the non-isothermal case, shown in Fig. \ref{F9}, it is can be observed that for $t \leq 3.8$, the total entropy production is primarily driven by NOEF, with ${\bar{\dot{S}}_{\text{NOEF}}}$ gradually decreasing. Between $3.8 < t \leq 5.6$, both ${\bar{\dot{S}}_{\text{NOMF}}}$ and ${\bar{\dot{S}}_{\text{NOEF}}}$ increase gradually, with similar amplitudes and growth rates. This suggests that during this period, the two entropy production mechanisms are competing and growing together. After $t = 5.6$, when the droplets fully cut-through, ${\bar{\dot{S}}_{\text{NOMF}}}$ surpasses ${\bar{\dot{S}}_{\text{NOEF}}}$ and begins to dominate the system's entropy production.

In summary, akin to the system's TNE effects, the two entropy production mechanisms associated with temperature and velocity gradients alternate in their dominance throughout the droplet coalescence process. During the damping oscillation phase,  ${\bar{\dot{S}}_{\text{NOMF}}}$, also exhibits noticeable damping oscillations.
%

Figure \ref{F10} displays the total entropy production of the system in both isothermal and non-isothermal conditions. As mentioned previously, in the isothermal case, entropy production is primarily due to NOMF; in the non-isothermal case, entropy production is contributed by both NOMF and NOEF, with the former being the dominant contributor. Compared to the isothermal case, the non-isothermal system exhibits a larger total entropy production due to the contribution of NOEF, though the entropy increase from NOEF is smaller, with ${S_{\text{NOEF}}}$ being approximately ${1}/{3}$ of ${S_{\text{NOMF}}}$. While the peak value of ${\bar{\dot{S}}_{\text{NOMF}}}$ in the isothermal case is about 2.6 times higher than in the non-isothermal case, the range of ${\dot{S}_{\text{NOMF}}}$ distribution is wider in the non-isothermal case, and its duration is longer. As a result, the entropy production from NOMF in the non-isothermal case, ${S_{\text{NOMF}}}$, is slightly greater than in the isothermal case.

\subsection{Effects of initial distance on the coalescence process}

\begin{figure*}[htbp]
\centering\includegraphics*
[width=1.\textwidth,trim=0.1 0.1 0.1 0.1,clip]{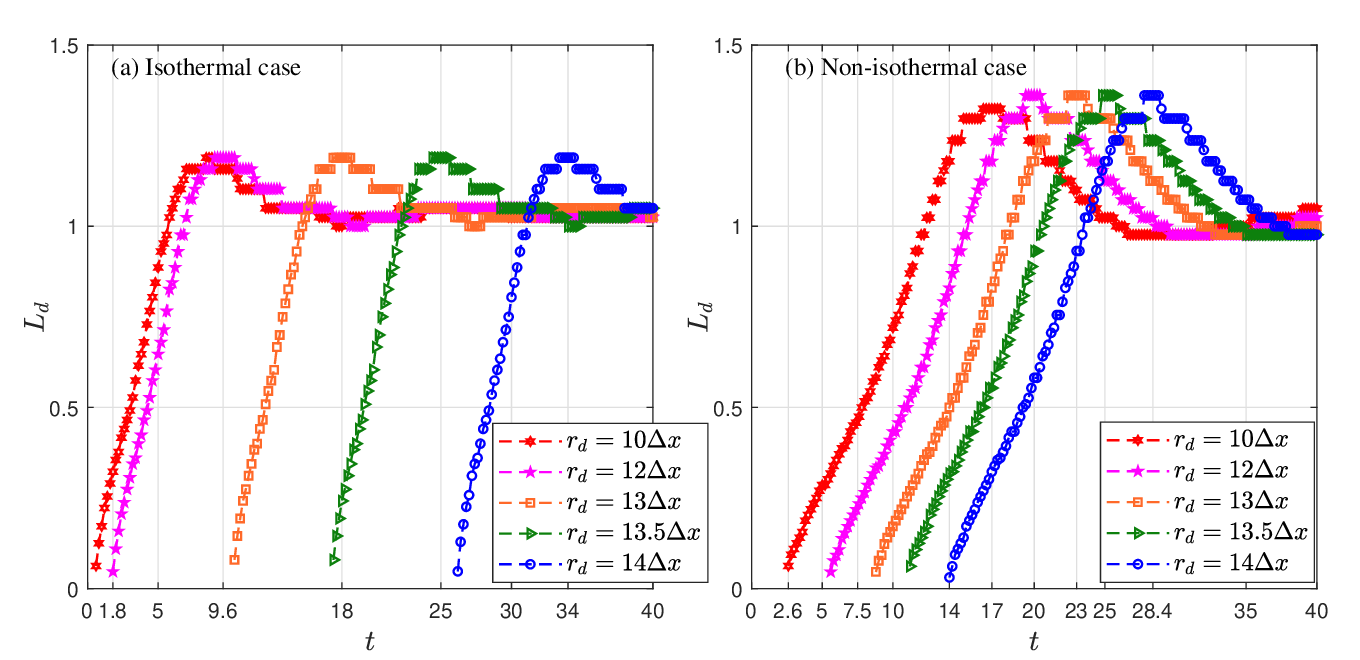}
 \caption{\centering{Effects of $r_d$ on the droplet's morphology factor $L_d$ under both isothermal and non-isothermal conditions.}}
 \label{F11}
\end{figure*}
\begin{figure}[htbp]
\centering\includegraphics*
[width=0.5\textwidth,trim=0.1 0.1 0.1 0.1,clip]{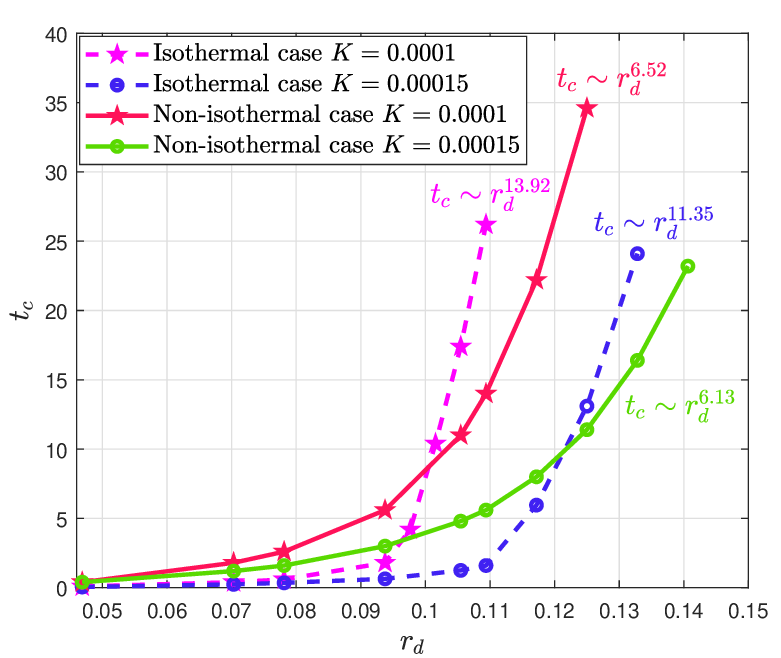}
 \caption{\centering{Effects of $r_d$ on the cut-through time $t_c$ under both isothermal and non-isothermal conditions, where $K=0.0001$ and $K=0.00015$.}}
 \label{F12}
\end{figure}
\begin{figure*}[htbp]
\centering\includegraphics*
[width=0.8\textwidth,trim=0.1 0.1 0.1 0.1,clip]{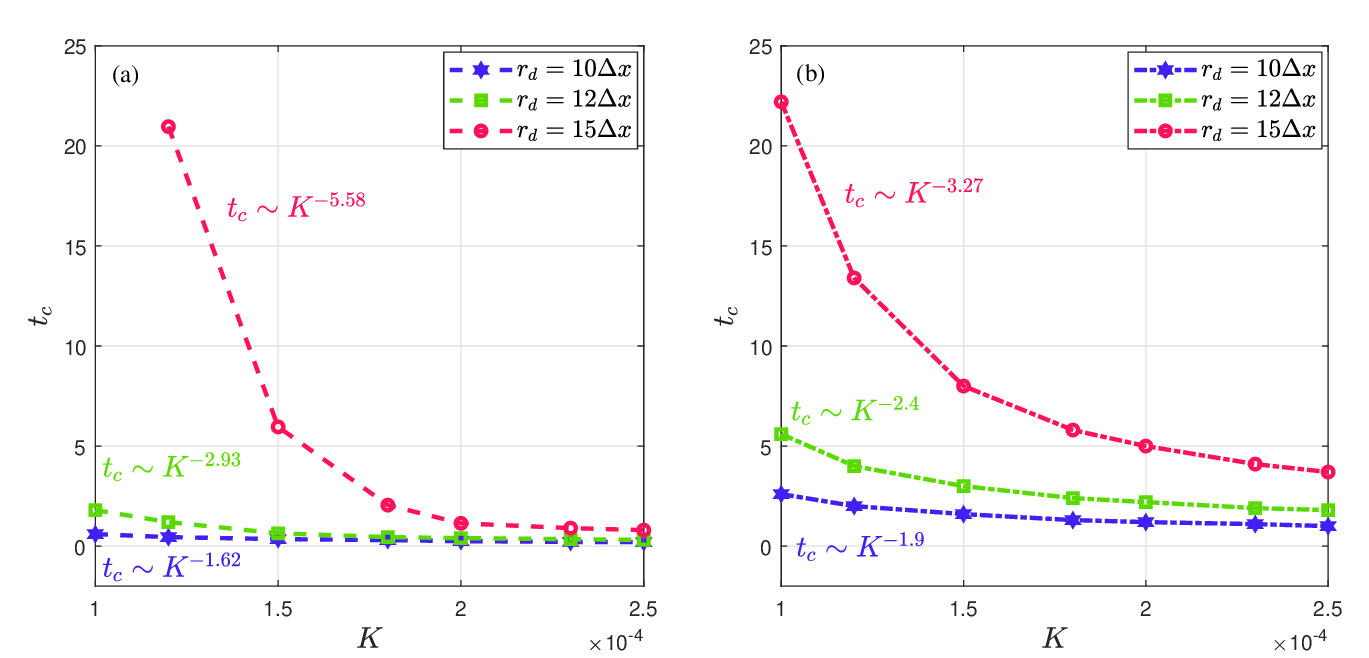}
 \caption{\centering{Effects of surface tension coefficient on the cut-through time $t_c$ under both isothermal and non-isothermal conditions.}}
 \label{F13}
\end{figure*}
The droplet coalescence process is influenced by various factors, with the initial distance between droplets playing a crucial role in coalescence velocity, mode, and final morphology. The initial separation between droplets determines the intensity of their interaction during coalescence, the formation and expansion of the liquid bridge, and the changes in surface tension and pressure gradients. For instance, in applications such as spraying and atomization, the initial distance between droplets significantly affects atomization efficiency, particle size, and spray uniformity. Optimizing this distance can improve atomization performance and enhance the effectiveness of processes like combustion, fertilizer spraying, and spray-based drug delivery  \cite{RN288, RN725, RN713}.

Figure \ref{F11} shows the evolution curves of the droplet's morphology factor \( L_d \) over time for different initial distances \( r_d \) between the droplets, under both isothermal and non-isothermal conditions. As shown, the evolution trends of \( L_d \) are similar for both isothermal and non-isothermal cases, except for the significant difference in the cut-through time \( t_c \). In the isothermal case, \( t_{l \rm max} - t_c \approx 7.8 \), while in the non-isothermal case, \( t_{l \rm max} - t_c \approx 14.4 \). This difference arises because after the cut-through, the rapid coalescence process is primarily controlled by surface tension. When droplet scale, surface tension coefficient, and viscosity remain constant, the droplet deformation, flow field distribution, and energy and momentum transfer are similar.

Figure \ref{F12} shows how the initial distance \(r_d\) between two droplets affects the cut-through time \(t_c\).
In both isothermal and non-isothermal cases, the cut-through time \(t_c\) increases with the initial droplet distance \(r_d\), following a power-law relationship.
For identical parameters, the exponent in the non-isothermal case is smaller than that in the isothermal case.
This behavior is due to the similarity in the system's dynamic.
Because, in the non-isothermal case, the temperature gradient induces energy dissipation mechanisms, such as the Marangoni effect and heat conduction, which delay the droplet coalescence stage.

When \(r_d\) is large, the contact mode between droplets changes in the isothermal case, reducing the effect of surface tension and significantly prolonging the droplet approach process, which sharply increases the cut-through time \(t_c\).
The functional relationships between \(t_c\) and \(r_d\) in the isothermal case are:
\[
{{t}_{c}}=6.45\times {{10}^{14}}r_{d}^{13.92}-0.093 \quad (K=0.0001)
\]
and
\[
{{t}_{c}}=2.18\times {{10}^{11}}r_{d}^{11.35}-0.043 \quad (K=0.00015).
\]
The functional relationships between \(t_c\) and \(r_d\) in the non-isothermal case are:
\[
{{t}_{c}}=7.35\times {{10}^{7}}r_{d}^{7.02}+0.964 \quad (K=0.0001)
\]
and
\[
{{t}_{c}}=3.72\times {{10}^{6}}r_{d}^{6.13}+0.826 \quad (K=0.00015).
\]
The power exponent decreases as the surface tension coefficient \(K\) increases.
In other words, a larger surface tension coefficient \(K\) results in a smaller crossing time \(t_c\).
As shown in Fig. \ref{F13}, \(t_c\) and \(K\) follow a negative power function relationship.
This can be attributed to two factors: First, surface tension is the primary driving force for droplet coalescence, regardless of whether the case is isothermal or non-isothermal. A larger surface tension coefficient \(K\) increases the surface tension, accelerating droplet approach and shortening the crossing time.
Second, a larger surface tension coefficient increases the liquid-vapor interface width  $w=\sqrt{{2K}/{({{{T}_{c}}}/{T}-1)}}$ \cite{RN779}, effectively reducing \(r_d\) and shortening the crossing time.

In the isothermal case shown in Fig. \ref{F13}, the specific functional relationships are:
\[
{{t}_{c}}=1.77\times {{10}^{-7}}{{K}^{-1.62}}+0.085 \quad (r_d=10\Delta x),
\]
\[
{{t}_{c}}=3.03\times {{10}^{-12}}{{K}^{-2.93}}+0.187 \quad (r_d=12\Delta x),
\]
and
\[
{{t}_{c}}=2.68\times {{10}^{-21}}{{K}^{-5.58}} \quad (r_d=15\Delta x).
\]
In the non-isothermal case shown in Fig. \ref{F13}, the specific functional relationships are:
\[
{{t}_{c}}=4.94\times {{10}^{-8}}{{K}^{-1.9}}+0.687 \quad (r_d=10\Delta x),
\]
\[
{{t}_{c}}=1.06\times {{10}^{-9}}{{K}^{-2.4}}+1.359 \quad (r_d=12\Delta x),
\]
and
\[
{{t}_{c}}=1.66\times {{10}^{-12}}{{K}^{-3.27}}+2.865 \quad (r_d=15\Delta x).
\]

\begin{figure*}[htbp]
\centering\includegraphics*
[width=1.\textwidth,trim=0.1 0.1 0.1 0.1,clip]{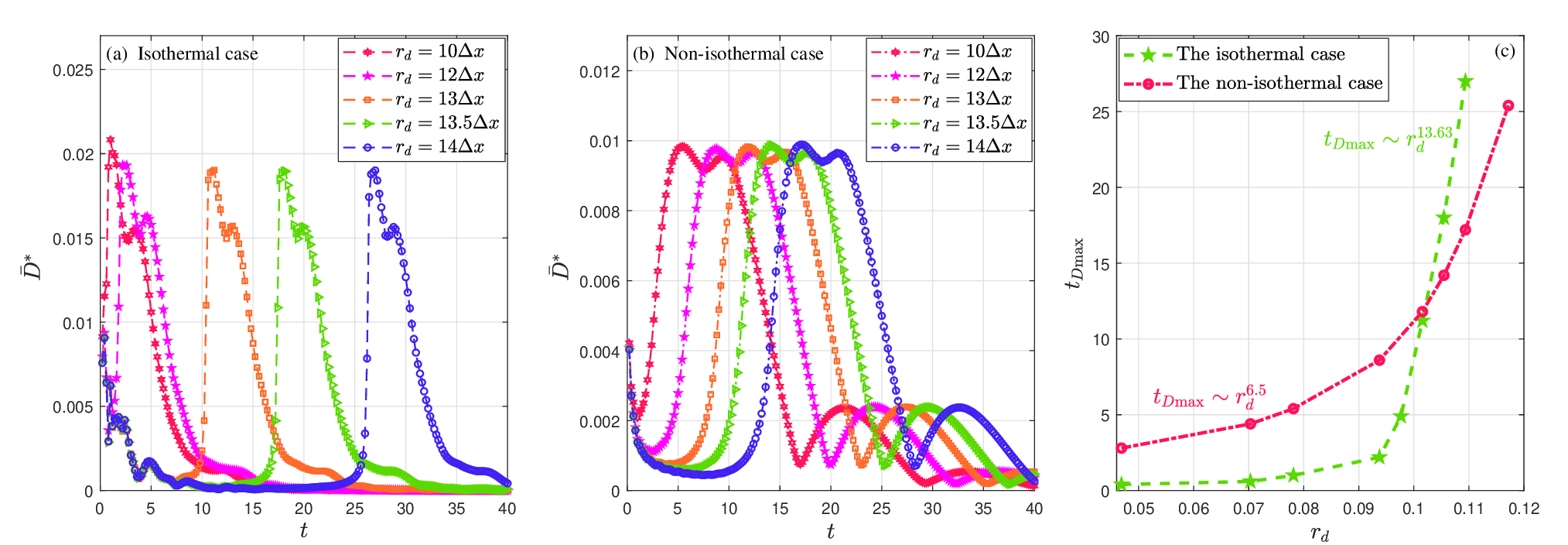}
 \caption{\centering{ {Effects of $r_d$ on  the total non-equilibrium strength $\bar D^{\ast}$ under both isothermal and non-isothermal conditions.}}}
 \label{F14}
\end{figure*}
\begin{figure*}[htbp]
\centering\includegraphics*
[width=1.\textwidth,trim=0.1 0.1 0.1 0.1,clip]{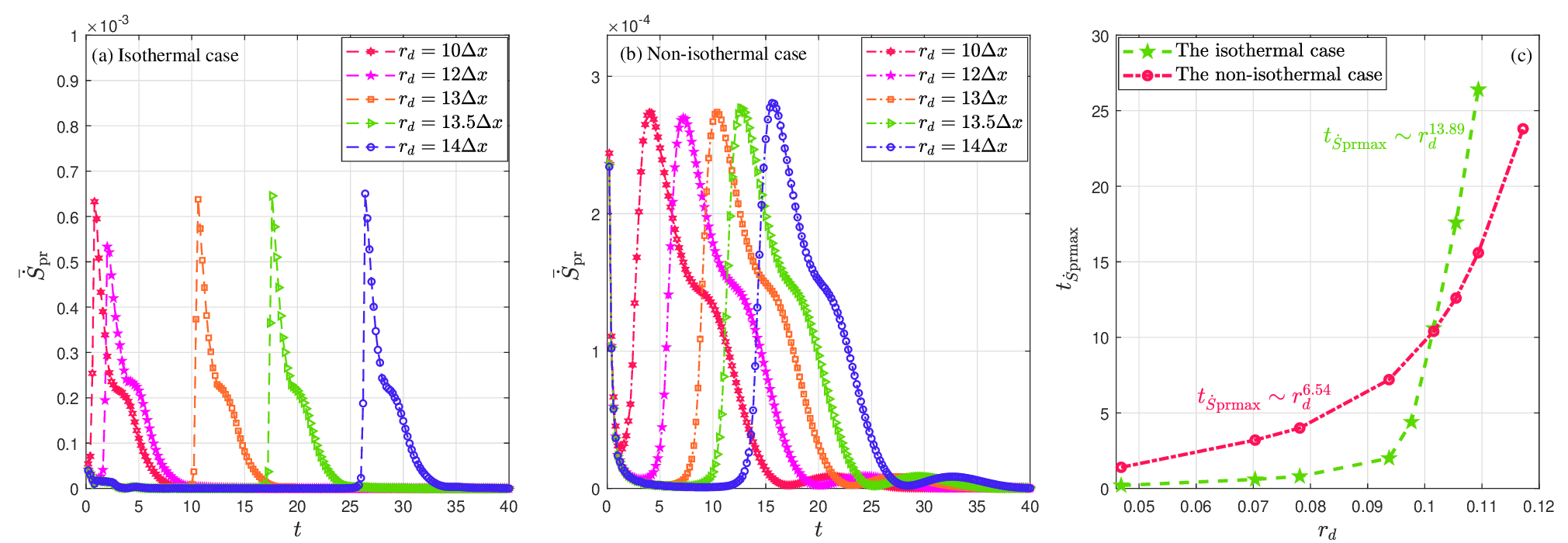}
 \caption{\centering{Effects of $r_d$ on  entropy production rate $\bar {\dot S}_{\rm pr}$ under both isothermal and non-isothermal conditions.}}
 \label{F15}
\end{figure*}

\begin{figure*}[htbp]
\centering\includegraphics*
[width=1.\textwidth,trim=0.1 0.1 0.1 0.1,clip]{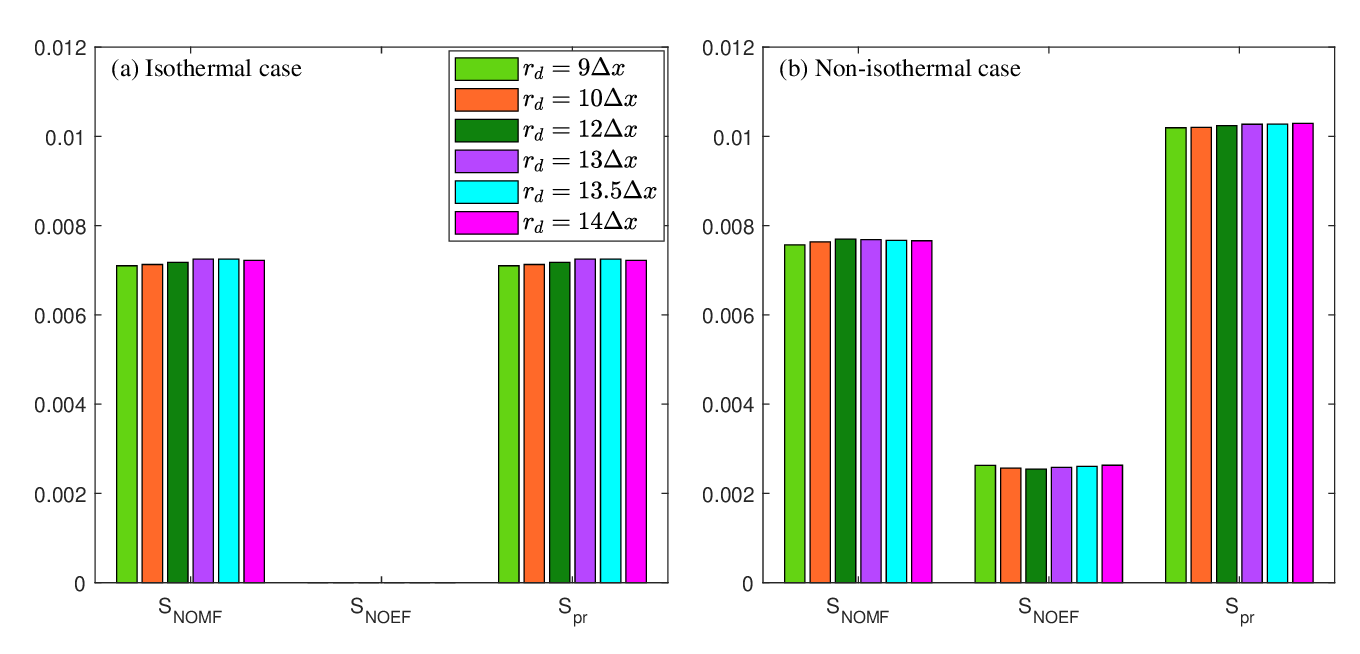}
 \caption{\centering{Effects of $r_d$ on total entropy production under both isothermal and non-isothermal conditions.}}
 \label{F16}
\end{figure*}

It should be noted that, in the isothermal case with $K=0.00001$, when $r_d=0.117$, the cut-through time $t_c$ approaches infinity, indicating that the droplets no longer merge. In fact, each set of parameters defines a limiting initial merging distance, $r_{d\rm max}$. For the same parameters, the limiting initial merging distance, $r_{d\rm max}$, is larger in the non-isothermal case than in the isothermal case. This is because, in the non-isothermal case, the temperature gradient-induced Marangoni effect and thermal convection increase the complexity of the flow field around the droplets, expanding the droplet coalescence range.

Figures \ref{F14}(a) and \ref{F14}(b) present the time evolution curves of the total non-equilibrium intensity, $\bar D^{\ast}$, for the isothermal and non-isothermal cases, respectively. As shown in Fig. \ref{F14}, the time evolution of $\bar D^{\ast}$, exhibit similarity for different $r_d$. In both cases, a power-law relationship exists between the time, $t_{D\rm max}$, at which the total non-equilibrium intensity reaches its maximum, and $r_d$. Moreover, the power exponent is greater in the isothermal case than in the non-isothermal case. The specific relationships between $t_{D\rm max}$ and $r_d$ are: ${{t}_{D\text{max}}}=3.46\times {{10}^{14}}{{r}^{13.63}_{d}}+0.205$ (isothermal case) and ${{t}_{D\text{max}}}=2.45\times {{10}^{7}}{{r}^{6.5}_{d}}+3.33$ (non-isothermal case).

Figures \ref{F15}(a) and \ref{F15}(b) illustrate the time evolution of the total entropy production rate, $\bar {\dot S}_{\rm pr}$, in the isothermal and non-isothermal cases, respectively. The initial distance, $r_d$, between the droplets has minimal impact on the time evolution and values of $\bar {\dot S}_{\rm pr}$, except that a larger $r_d$ delays the time at which $\bar {\dot S}_{\rm pr}$ reaches its maximum. Both isothermal and non-isothermal cases also exhibit power-law relationships between $t_{\dot S \rm pr max}$ (when $\bar {\dot S}_{\rm pr}$ reaches its maximum) and $r_d$, with the exponent in the isothermal case larger than that in the non-isothermal case. Additionally, $r_d$ has little influence on the amounts of the three types of entropy production ($S_{\rm NOMF}$, $S_{\rm NOEF}$, $S_{\rm pr}$), as shown in Fig. \ref{F16}. The relationships between $t_{\dot S \rm pr max}$ and $r_d$ in Fig. \ref{F15}(c) also follow power-laws: ${{t}_{\dot{S}\text{prmax}}}=5.99 \times 10^{14} r_d^{13.89} + 0.077$ (isothermal) and ${{t}_{\dot{S}\text{prmax}}}=2.63 \times 10^{7} r_d^{6.54} + 2.027$ (non-isothermal). 

\section{Conclusions and outlooks}\label{Sec4}

This paper presents a detailed analysis of the coalescence process of two initially static droplets under both isothermal and non-isothermal conditions, focusing on the impact of the initial distance between the droplets. Using the DBM method, the study explores several key aspects.
The main findings include:

(1) Coalescence dynamics: In both cases, surface tension drives droplet coalescence, while the resistance arises from an isotropic pressure gradient. However, in the non-isothermal case, the temperature rise due to phase transition latent heat significantly increases the isotropic pressure gradient, though it slightly increases the surface tension gradient near the droplet contact point. This results in a smaller total pressure tensor gradient leading to the deferment of droplet coalescence in the non-isothermal case.

(2) Flow field complexity: The non-isothermal case exhibits a significantly more complex flow field due to the effects of thermal Marangoni forces and heat conduction, which contribute to greater complexity in the evolution of droplet morphology. The spatial distributions of various non-equilibrium driving forces also differ markedly between the two cases.

(3) Thermodynamic non-equilibrium: In the isothermal case, non-equilibrium behaviors are primarily driven by velocity gradients, with a stronger non-equilibrium intensity. In the non-isothermal case, however, non-equilibrium behaviors alternate between temperature gradients and velocity gradients as the droplets begin to merge, leading to more complex oscillations in non-equilibrium and entropy production characteristics. The shift in the dominant factor reflects the complexity of the system, which arises from the combined influence of multiple driving factors.

(4) Entropy production: In the isothermal case, entropy production is solely due to NOMF, while in the non-isothermal case, both NOMF and NOEF contribute, with NOMF being the dominant contributor. The contributions from NOMF and NOEF alternate before and after droplet cut-through.

(5) Effect of initial distance: Increasing the initial distance between the droplets delays the coalescence time ($t_c$) but does not significantly alter the system's morphological, non-equilibrium, or entropy production characteristics. The relationships between key times (e.g., $t_c$, $t_{D\rm max}$, $t_{\dot S_{\rm prmax}}$) and $r_d$ follow a power-law, with larger exponents in the isothermal case.

This study demonstrates that the dynamics of droplet coalescence are heavily influenced by thermodynamic non-equilibrium effects, particularly in the non-isothermal case.  {Non-equilibrium measures can provides an earlier and more sensitive means of detecting the dominant processes associated with different mechanisms and their transitions.}
The initial distance between droplets primarily impacts the timing of coalescence, with minimal effect on other characteristics, and the system exhibits similar behavior in its evolution.

 {In future work, the following three research directions are proposed to expand and deepen the current study:}

 {(1) Extension beyond the local equilibrium assumption: The current external force are introduced base on the equilibrium distribution function. This assumption requires that the phase separation process proceeds much slower than the relaxation of TNE effects, i.e., \( t_{\rm ps} \gg \tau \), where \( t_{\rm ps} \) denotes the characteristic time of phase separation.
However, as the phase separation process accelerates and the degree of discreteness or nonequilibrium increases, the validity of the local equilibrium assumption becomes questionable.
In particular, when the equation of state becomes invalid or unknown, the current modeling approach becomes inadequate.
Under such conditions, it becomes necessary to explicitly incorporate intermolecular potential functions to accurately describe molecular interactions and recover the correct thermodynamic behavior \cite{LH2014,LH2018}.}

 {(2) Influence of solid walls and shear flows: In microfluidic systems, the presence of solid walls and external shear flows has a significant impact on the droplet coalescence process.
Specifically, factors such as the contact angle, wall temperature, and shear strength play critical roles in determining the dynamics of droplet deformation, coalescence, and stabilization.
Future investigations should incorporate wetting boundary conditions and shear effects to better capture the complex physical phenomena occurring in confined or structured environments.}

 {(3)  Droplet deformation and breakup under shock waves:} Future research will focus on the deformation, fragmentation, cavitation, and jetting processes of droplets under shock waves \cite{RN784,RN828,RN790}—phenomena with far-reaching applications. In areas like spray combustion, engine injection, and aerospace, the shape and breakup of droplets are crucial in determining combustion efficiency, thrust, and dynamic performance. From a physical standpoint, this process is marked by increasingly significant thermodynamic non-equilibrium effects, making it an essential domain where kinetic methods are necessary for precise modeling and analysis.

\begin{acknowledgments}
This work was supported by the National Natural Science Foundation of China (Grant Nos.  52278119, 11875001 and 12172061), the Hebei Outstanding Youth Science Foundation (Grant No. A2023409003), the Central Guidance on Local Science and Technology Development Fund of Hebei Province (Grant No. 226Z7601G), Science Foundation of NCIAE (Grant No. ZD-2025-06), Science Research Project of Hebei Education Department (Grant No. QN2025184), and the Foundation of the National Key Laboratory of Shock Wave and Detonation Physics (Grant No. JCKYS2023212003).
\end{acknowledgments}

\section*{Appendix}
 {Table \ref{TT} shows definitions, components, and their physical meanings of TNE effects:}

 \renewcommand{\arraystretch}{1.5} 
\begin{table*}[htbp]
\centering
\caption{Definitions, components, and physical interpretations of TNE quantities in DBM.}
\begin{tabular}{m{1cm}<{\centering} m{6.5cm}<{\centering} m{4cm}<{\centering} m{5cm}<{\centering}}
\hline
\hline
Quantity & Definition & Components & Physical Meaning \\
\hline
$\bm{\Delta}^{ *}_{2}$ & $\sum_{i}(f_{i}-f^{eq}_{i})\mathbf{v}^{*}_{i}\mathbf{v}^{*}_{i}$ & $xx$, $xy$, $yy$ & non-organized momentum flux (NOMF) \\
\hline
$\bm{\Delta}^{ *}_{3,1}$ & $\frac{1}{2}\sum_{i}(f_{i}-f^{eq}_{i})(\mathbf{v}^{*}_{i}\cdot\mathbf{v}^{*}_{i}+\eta_{i}^{2})\mathbf{v}^{*}_{i}$ & $x$, $y$ & non-organized energy flux (NOEF) \\
\hline
$\bm{\Delta}^{*}_{3}$ & $\sum_{i}(f_{i}-f^{eq}_{i})\mathbf{v}^{*}_{i}\mathbf{v}^{*}_{i}\mathbf{v}^{*}_{i}$ & $xxx$, $xxy$, $xyy$, $yyy$ & flux of NOMF \\
\hline
$\bm{\Delta}^{*}_{4,2}$ & $\frac{1}{2}\sum_{i}(f_{i}-f^{eq}_{i})(\mathbf{v}^{*}_{i}\cdot\mathbf{v}^{*}_{i}+\eta_{i}^{2})\mathbf{v}^{*}_{i}\mathbf{v}^{*}_{i}$ & $xx$, $xy$, $yy$ & flux of NOEF \\
\hline
$\bar D_{2}^{*}$ & $\sum{\rho\left| \bm{\Delta}^{ *2}_{2} \right|}/{\rho}$ & scalar & density-weighted average NOMF intensity \\
\hline
$\bar D_{3}^{*}$ & $\sum{\rho \left| \bm{\Delta}^{*2}_{3} \right|}/{\rho}$ & scalar & density-weighted average of the flux of NOMF intensity \\
\hline
$\bar D_{3,1}^{*}$ & $\sum{\rho \left| \bm{\Delta}^{*2}_{3,1} \right|}/{\rho}$ & scalar & density-weighted average NOEF intensity \\
\hline
$\bar D_{4,2}^{*}$ & $\sum{\rho \left| \bm{\Delta}^{ *2}_{4,2} \right|}/{\rho}$ & scalar & density-weighted average of the flux of NOEF intensity \\
\hline
$\bar D^{*}$ & $\sum{\rho \sqrt{ \left| {\bm{\Delta}^{*2}_{2}} \right| + \left| {\bm{\Delta}^{*2}_{3,1}} \right| + \left| {\bm{\Delta}^{*2}_{3}}\right| + \left| {\bm{\Delta}^{*2}_{4,2}}\right|}/{\rho}}$ & scalar & density-weighted average global TNE intensity \\
\hline
\hline
\end{tabular}
\label{TT}
\end{table*}


\section*{Data Availability}
The data that support the findings of this study are available from the corresponding author upon reasonable request.

\section*{References}
\bibliography{Droplet-Coalescence-Kinetic}

\end{document}